\documentclass[journal]{IEEEtran}

\ifCLASSINFOpdf
\else
    \usepackage[dvips]{graphicx}
\fi
\usepackage{url}

\usepackage{graphicx}

\usepackage{cite}
\usepackage{amsmath}
\usepackage{amssymb}
\usepackage{hyperref}
\usepackage{graphicx}
\usepackage{algorithm}
\usepackage{algpseudocode}
\usepackage{verbatim}
\usepackage{bbm} 
\usepackage{pdfpages}
\usepackage{braket}
\usepackage{mathrsfs}
\usepackage[caption=false]{subfig}
\usepackage{tikz}
\usetikzlibrary{positioning}
\usepackage{multirow}
\usepackage{booktabs} 

\newtheorem{lemma}{\textbf{Lemma}}

\newtheorem{defn}{\textbf{Definition}}
\newtheorem{prop}{\textbf{Proposition}}
\newtheorem{remark}{\textbf{Remark}}
\newtheorem{prob}{\textbf{Problem}}

\floatname{algorithm}{Algorithm}

\algdef{SE}[SUBALG]{Indent}{EndIndent}{}{\algorithmicend\ }%
\algtext*{Indent}
\algtext*{EndIndent}
\newcommand{\qed}{\hfill\blacksquare}
\DeclareMathOperator{\diag}{diag}

\newcommand{\Exp}{\mathbb{E}}
\newcommand{\red}[1]{\textcolor{red}{#1}}

\begin{document}

\title{Phase-Only Zero-Forcing for Secure Wireless Communication in Multi-User Systems}

\author{Jordan Hong, \IEEEmembership{Student Member, IEEE}, Safwan Jamal, \IEEEmembership{Student Member, IEEE}, and Ashish Khisti, \IEEEmembership{Member, IEEE}}

\maketitle

\begin{abstract}
Artificial noise (AN) transmission is a physical layer security technique in multi-antenna wireless communication systems.  
Synthetic noise is broadcast to all receivers except designated legitimate users via beamforming in the legitimate users' null space.
We consider AN transmission employing a single RF chain and analog beamforming, where beamforming vectors maintain constant magnitude while allowing arbitrary phases.

Our primary objective is to design a constant-magnitude vector capable of nullifying multiple users' channel vectors simultaneously. 
To tackle this zero-forcing problem, we propose a novel successive partition zero-forcing (SPZF) scheme, which transforms the multi-user zero-forcing task into optimizing channel partitioning to minimize outage probability.

The SPZF scheme can be generalized to any number of users, but our analysis focuses on the two-user case.
Theoretical analysis reveals that our proposed SPZF scheme can attain arbitrarily low outage probability in the limit of a large number of transmit antennas.
We present three partition algorithms (random, iterative, and genetic) to minimize the outage probability. 
The outage probabilities and secrecy rates of the three partition algorithms are compared via numerical simulations. 
We find that the more advanced partition algorithms (iterative and genetic) achieve higher secrecy rates than the random algorithm, particularly under conditions of high signal-to-noise ratio (SNR), large number of eavesdroppers, or small number of transmit antennas.
\end{abstract}


\IEEEpeerreviewmaketitle

\section{Introduction}

\IEEEPARstart{M}{assive} MIMO (large-scale antenna array systems) is a key enabler for next-generation wireless networks (5G and beyond) because of its high spectral efficiency, high throughput, and energy efficiency \cite{andrews_2014}, \cite{larsson_2014}. The use of large-scale arrays is enabled by millimeter-wave (mmWave) technologies, which are characterized by the high frequency (30 - 300 GHz) range and small wavelengths in the millimeter scale. The small wavelengths permit smaller antenna space and hence more antenna arrays. In addition, massive MIMO can be built using inexpensive and low-power antennas with limited RF (radio frequency) chains, which is attractive to future generation wireless networks that are increasing in number and diversity of Internet of Things (IoT) devices.

\par Alongside these advancements, the importance of robust security in wireless networks has become increasingly evident. Toward this end, the application of Artificial Noise (AN) transmission as a security measure is gaining prominence in massive MIMO networks. AN transmission is a physical layer security technique in which information symbols and synthetic noise symbols are transmitted simultaneously through separate sets of antenna arrays \cite{zhao_lee_khisti_2016}. The synthetic noise symbols are constructed to lie in the null space of the legitimate receivers. As such, the legitimate user's signal is not interfered with by the artificial noise, but all other eavesdropping users observe interference.

\par We consider a massive MIMO network with \emph{analog beamforming}, where each RF chain is connected to a set of phase shifters that control the phase of each antenna. This requires all elements in the beamforming vector to have constant magnitudes while the phases can be arbitrary. Under this non-convex constraint, we aim to design a phase-only beamforming algorithm that extends to multiple users. The main contributions of this paper are:
\begin{itemize}
    \item We propose a novel successive partition zero-forcing (SPZF) approach to find a set of phases that zero-forces multiple complex channel vectors. We show that the SPZF approach is capable of achieving an arbitrarily low outage probability. 
    \item We transform the zero-forcing problem to a partition problem and design three partition algorithms (random, iterative, genetic). We compare the outage probability, secrecy rates, and runtime of the three algorithms using numerical simulations for the two-user scenario, using the i.i.d. Rayleigh fading channel model and the geometric multipath channel model.
\end{itemize}

\par The remainder of the paper is organized as follows. The system model is given in Section \ref{sec:system-model}. The research problem is formulated in Section \ref{sec:problem-formulation}. In Section \ref{sec:spzf}, we propose the SPZF scheme and formulate the research problem into a channel partition problem. In Section \ref{sec:proposed-partition}, we propose three partition algorithms with varying performance and complexities. In Section \ref{sec:simulation-results}, we present simulation results of the beamforming outage probability and secrecy rate. Section \ref{sec:conclusion} concludes the paper.

\par \emph{Notation:} We use bold upper-case and lower-case letters to represent matrices and vectors, respectively. Underlined lower-case letters also denote a vector. $j \triangleq \sqrt{-1}$. $(\cdot)^T$ denotes the transpose. $| \cdot |$ denotes the absolute value. $\mathbb{C}$ denotes the set of complex numbers. $\Pr(\cdot)$ denotes the probability of the input event. $\mathcal{CN}(u, \sigma^2)$ represents the circularly symmetric complex Gaussian distribution with mean $u$ and variance $\sigma^2$. $\Exp{[\cdot]}$ denotes the expected value of a random variable. $\mathbbm{1}[\cdot]$ denotes the indicator function. $\mathcal{H}(x)$ is the Heaviside unit step function, i.e. $\mathcal{H}(x) = \mathbbm{1}[x > 0]$.

\section{System Model}
\label{sec:system-model}
\par We consider a MU-MISOME (Multi-user, multi-input, single-output, multi-eavesdropper) system \cite{khisti_wornell_2010}. The downlink system configuration is comprised of:
\begin{enumerate}
    \item One transmitter with $N$ antenna. The transmitter generates an input vector $\mathbf{x} \in \mathbb{C}^N$ under the average power constraint $P$, i.e. $\Exp[\, \| x \|^2\, ] \leq P$.
    \item $K$ legitimate receivers, each with a single antenna. We use the subscripts $k \in [1:K]$ to represent the $K$ legitimate receivers.
    \item One eavesdropper with $N_e$ antennas.
\end{enumerate}

\par At discrete time $t$, the signals received by the $k$-th legitimate receiver and the  eavesdropper are given as follows: \begin{equation}
    y_k (t) = \mathbf{h_k^T x}(t) + n_k (t),
    \label{eq:k-receiver-signal}
\end{equation}
\begin{equation}
    \mathbf{z}(t) = \mathbf{Gx} (t) + \mathbf{n}_e (t),
    \label{eq:e-eve-signal}
\end{equation}
where $\mathbf{x}(t)$ is the symbol vector (precoded symbol) at time $t$ that is transmitted from the transmitter, $\mathbf{h_k} \in \mathbb{C} ^N, \mathbf{G} \in \mathbb{C} ^{N_e \times N}$ are the channel vector and matrix associated with the $k$-th receiver and the eavesdropper respectively. $n_k \sim \mathcal{CN}(0,1)$ and $\mathbf{n}_e \sim \mathcal{CN}(0,\mathbf{I})$ are the additive white Gaussian noise (AWGN) of the channel for the $k$-th receiver and the eavesdropper.

\begin{figure}[htpb]
    \centering
    \includegraphics[width=0.5\textwidth]{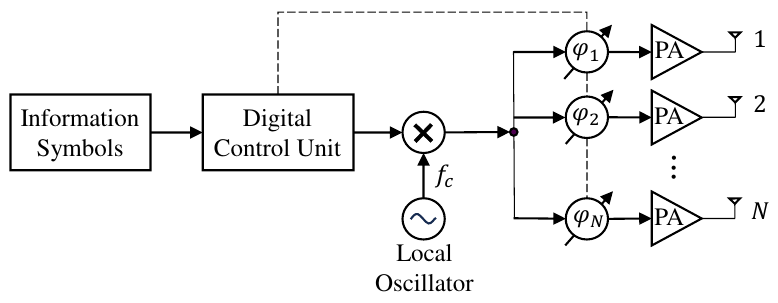}
    \caption{Schematics of the phased-array transmission structure}
    \label{fig:phase-array}
\end{figure}

\par Let the channel vector and matrix take the form of
\begin{align}
                \mathbf{h_k} &= [h_{k1},\cdots , h_{kN}]^T, \\
                \mathbf{G} &= [\mathbf{g_1}, \cdots, \mathbf{g}_{N_e} ]^T, \\
                \mathbf{g_e} &= [g_{e1}, \cdots , g_{eN}]^T, \quad e \in [1:N_e],
\end{align}
\par We consider the following two channel models for $\mathbf{h_k}$ and $\mathbf{g_e}$:
\begin{enumerate}
    \item i.i.d. Rayleigh fading model: $h_{ki} \sim \mathcal{CN}(0,\sigma^2)$ for $k \in [1:K]$ and $i \in [1:N]$, $g_{ei} \sim \mathcal{CN}(0,\sigma^2)$ for $e \in [1:N_e]$ and $i \in [1:N]$ for some $\sigma^2 > 0$.
    \item Geometric model with $L$ paths:
        \begin{equation}
            \mathbf{h_k} = \sqrt{\frac{1}{L}} \sum_{l=1}^{L} \alpha_l \mathbf{a_l}^k (\phi_l^k),
            \label{eq:geometric-channel-model}
        \end{equation}
        for $k = [1:K]$, where $\alpha_l \sim \mathcal{CN}(0, 1)$ denotes the channel gain for the $l$-th path and $\mathbf{a_l}^k (\phi_l^k)$ denotes the antenna array response vector given as follows assuming that the transmitter's antennas form a uniform linear array (ULA):
        \begin{equation}
            \mathbf{a_l}^k (\phi_l^k) = \frac{1}{\sqrt{N}}[1, e^{j \frac{2\pi}{\lambda} d \sin \phi_l^k}, \dots, e^{j \frac{2\pi}{\lambda} (N-1) d \sin \phi_l^k}]^T,
            \label{eq:geometric-channel-model-a}
        \end{equation}
        where $\phi_l^k$ denotes the random azimuth angle of departure for the $l$-path to the $k$-th receiver, which is drawn independently and uniformly over $[0, 2 \pi]$; $d$ denotes the antenna spacing, $\lambda$ denotes the wavelength. Each $\mathbf{g_e}$ in $\mathbf{G}$ follows the same model in Equation \ref{eq:geometric-channel-model}. In the numerical simulations of this paper, we set $d = \frac{\lambda}{2}$ as suggested by \cite{chu1997semi}.
\end{enumerate}

\par Figure \ref{fig:phase-array} illustrates the system model. The transmitter uses a single RF chain for AN transmission. The precoded symbol vector $\mathbf{x}$ takes the following form:
\begin{equation}
    \mathbf{x} = \sqrt{\frac{P}{N}} s \cdot \mathbf{w},
    \label{eq:precoded-symbol}
\end{equation}
where $s \sim \mathcal{CN}(0,1)$ is the noise symbol being transmitted, and $\mathbf{w}$ is the beamforming vector. In this study, we focus on an analog beamforming architecture \cite{sohrabi_yu_2016}, implemented by a single analog phase shifter at each antenna. This architecture requires $\mathbf{w}$ to have the form:
\begin{equation}
    \mathbf{w} = [e^{j \varphi_1}, \cdots , e^{j \varphi_N} ]^T.
    \label{eq:beamform-w}
\end{equation}

\par From Equation \ref{eq:k-receiver-signal}, if at discrete time $t$ the transmitter is sending an artificially constructed noise symbol $s(t)$, then the received symbol at time $t$ is:
    \begin{equation}
        y_k (t) = \mathbf{h_k^T w} \sqrt{\frac{P}{N}} s + n(t).
        \label{eq:received-noise-signal}
    \end{equation}
    In the context of AN transmission, for the legitimate receiver to be agnostic to the artificial noise, the beamforming vector $\mathbf{w}$ should zero-force $\mathbf{h_k^Tw}$ in Equation \ref{eq:received-noise-signal}. We refer to a feasible $\mathbf{w}$ as a \emph{phase-only zero-forcing} (PZF) vector.

    \par The objective of this research is to design a beamforming algorithm that solves for a common beamforming vector $\mathbf{w}$ that zero-forces all channel vectors in a multiuser setting, while adhering to the constant modulus constraint (Equation \ref{eq:beamform-w}). Specifically, we aim to find a set of phases ($\varphi_i$, $i = 1, \dots, N$) (in Equation \ref{eq:beamform-w}), such that the beamforming vector $\mathbf{w}$ lies in the null space of all $K$ legitimate receivers' channels, i.e.:
    \begin{equation}
        \mathbf{h_k^T w} = \sum_{i=1}^{N} h_{ki} e^{j \varphi_i} = 0 \quad \forall k = 1, \dots, K.
        \label{eq:pzf}
    \end{equation}

\section{Problem Formation and Related Work}
\label{sec:problem-formulation}
\par In this section, we consider three variants of the zero-forcing problem with respect to the number of legitimate receivers ($K$): a) general case of $K \geq 1$, b) special case of $K = 1$, and c) special case of $K = 2$. Existing research that pertains to the problems will be surveyed in each subsection.

\subsection{Phase-only zero-forcing: the general $K$ case}
\par The task of general case of $K$ users is to design $\mathbf{w}$ (\ref{eq:beamform-w}) to satisfy Equation \ref{eq:pzf}. 
A numerical algorithm for the same system model has been proposed in \cite{zhao_lee_khisti_2016}.
The authors reduced the problem to an unconstrained non-linear programming problem and subsequently solved the non-linear programming problem using the Gaussian-Newton method. They showed that the non-linear programming approach performed significantly better than the heuristic relaxation approach. However, the non-linear programming numerical algorithm suffers from the drawback of high complexity.

A slight variation of the problem has been solved in \cite{9690054}. Their system model is identical to this study except that they considered a dual-phase shifter. This results in a relaxed constraint of the beamforming vector $\mathbf{w}$; the unit modulus constraint (Equation \ref{eq:beamform-w}) is replaced by a continuous interval. This relaxation significantly changes the problem.

\subsection{Phase-only zero-forcing: special case of $K = 1$}
\par The special case of a single legitimate receiver has been solved in \cite{zhao_lee_khisti_2016} and \cite{zhang_xia_he_zhang_2020}. For brevity, in the single legitimate receiver case, we denote the channel vector $\mathbf{h_1}$ as $\mathbf{h} = [h_1,\cdots , h_N]^T$. The zero-forcing condition in Equation \ref{eq:pzf} then becomes:
\begin{equation}
    \mathbf{h^T w} = \sum_{i=1}^{N} h_i e^{j \varphi_i} = 0.
    \label{eq:pzf-K1}
\end{equation}

\par
\begin{figure}[htpb]
        \centering        \includegraphics[width=0.5\textwidth]{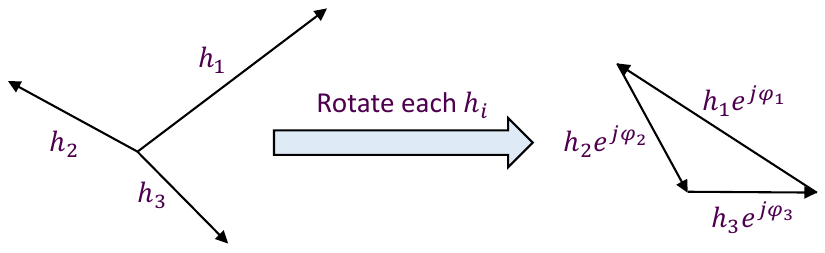}
        \caption{Geometric interpretation of Equation \ref{eq:pzf-K1} when N = 3: rotating each $h_i$ to form a triangle.}
        \label{fig:triangle-inequality}
    \end{figure}

 A natural geometric interpretation of the phase-only zero-forcing condition of Equation \ref{eq:pzf-K1} has been established in \cite{zhang_xia_he_zhang_2020}. In the complex plane, $h_i \in \mathbb{C}$ can be viewed as 2-dimensional vectors with real ($\mathfrak{R}\{h_i\}$) and imaginary ($\mathfrak{T}\{h_i\}$) components. Further, $\mathbf{w}$ rotates each vector $h_i \in \mathbf{h}$ by an arbitrarily chosen angle $\varphi_i$. Thus, Equation \ref{eq:pzf-K1} reduces to the problem of rotating $N$ vectors in such a way that their sum is zero. Geometrically, this is equivalent to rotating a set of 2-dimensional vectors in the complex plane to form a polygon. Figure \ref{fig:triangle-inequality} shows an example with $N = 3$.

    \par The following important lemma (proven in \cite{zhao_lee_khisti_2016}) provides a necessary and sufficient condition for a qualifying beamforming vector $\mathbf{w}$ to exist. Note that this is a generalization of the condition to complete a triangle (where $N = 3$) to polygons ($N \geq 3$). Henceforth, we will refer to this condition as the polygon inequality.
    \begin{lemma}[Polygon inequality]
        \label{lem:pzf-condition}
        Given any $\mathbf{h} \in \mathbb{C} ^N$ with $N \geq 3$, there exists a PZF vector $\mathbf{w}$ that satisfies Equation \ref{eq:pzf} if and only if:
        \begin{equation}
            \max_{1 \le i \le N} |h_i| \leq \frac{1}{2}\sum_{i=1}^{N} |h_i|.
            \label{eq:pzf-polygon-inequality}
        \end{equation}
    \end{lemma}

    \par The following lemma (presented in \cite{zhao_lee_khisti_2016}) shows that for the i.i.d. Rayleigh fading channel, the probability of failing to meet the polygon inequality decreases exponentially in $N^2$ (where $N$ is the number of transmit antennas).
    \begin{lemma}
        Let $\varepsilon$ denote the set of channel vectors for which a PZF vector does not exist, i.e.
        \begin{equation}
            \varepsilon \triangleq \{ \mathbf{h} \in \mathbb{C}^N: \max_{1 \le i \le N} | h_i| > \frac{1}{2}\sum_{i=1}^{N} |h_i| \}.
        \end{equation}
        If entries of $\mathbf{h}$ are i.i.d. $\mathcal{CN}(0,\sigma^2)$, then
        \begin{equation}
            \Pr(\varepsilon) \approx N e^{- \frac{N^2 \pi}{16}}, N \gg 1.
            \label{eq:pzf-wlln-bound}
        \end{equation}
        \label{lem:pzf-err-prob}
    \end{lemma}

    A polygon construction algorithm for the single user case was presented in \cite{zhang_xia_he_zhang_2020}. The authors leveraged the geometric intuition of completing a polygon and devised a phase solver that guarantees a zero-forcing solution $\mathbf{w}$ for Equation \ref{eq:pzf} provided that the polygon inequality (Lemma \ref{lem:pzf-condition}) is satisfied. However, the polygon construction method cannot be easily extended to zero-forcing multiple user channels simultaneously. In this study, we aim to build on the polygon construction method to design a multi-user zero-forcing algorithm.

\subsection{Phase-only zero-forcing for the special case of $K = 2$}
\par To tackle the problem of general $K$, we first analyze the special case of two legitimate receivers ($K = 2$). 
We consider two independent i.i.d. Rayleigh channels $\mathbf{h_1}$ and $\mathbf{h_2}$ and aim to solve for $\mathbf{w}$ that satisfies:

\begin{equation}
    \begin{bmatrix} \mathbf{h_1^T} \\ \mathbf{h_2^T} \end{bmatrix} \mathbf{w}
    = \sum_{i=1}^{N} \begin{bmatrix} h_{1i} \\ h_{2i} \end{bmatrix} e^{j \varphi_i} = 0.
    \label{eq:pzfh1h2}
\end{equation}

\par From a geometric perspective, the problem is equivalent to solving a set of phase rotations that complete two polygons simultaneously. To the best of our knowledge, there has been no analytical method to complete two polygons simultaneously; there are only numerical algorithms reported in \cite{zhao_lee_khisti_2016} for the aforementioned case of general $K$ users.

\section{Successive Partition Zero-forcing}
\label{sec:spzf}
\par We propose a novel successive partition zero-forcing (SPZF) approach for the special case of $K=2$ and show that this algorithm can be easily extended to general $K \geq 2$. To the best of our knowledge, this successive zero-forcing approach has never been reported. Next, we analyze the SPZF approach for the special case of $K = 2$ in the i.i.d. Rayleigh fading channel and show that the problem can be viewed as a problem of constructing a partition to minimize the outage probability. An outage occurs when a solution for  $\mathbf{w}$ cannot be found. Since the general outage probability does not have a closed mathematical form, we consider a more constrained problem that fixes the number of partition sets at exactly $m$. We construct and minimize an upper bound for the outage probability for the problem of fixed $m$.

The intuition of the SPZF approach system is to decouple a system of zero-forcing equations and solve (zero-force) one equation at each step. 
In each step, the equation is partitioned into multiple sets, where each partition set is individually zero-forced. 
For simplicity, we explain the algorithm for the $K = 2$ case. Generalization to $K > 2$ can be done recursively, provided that $N \geq 3^K$; we show this in Appendix \ref{app:spzf-generalization}.


\subsection{Motivational Example for the special case of $K = 2$} \label{sse:spzf-ex}
The following example illustrates the key idea of SPZF. We consider Equation \ref{eq:pzfh1h2} with $N = 9$ and partition the system of equations into $m = 3$ subsets.
\begin{equation}
    \sum_{i = 1}^{3} \begin{bmatrix} h_{1i} \\ h_{2i}\end{bmatrix} e^{j \varphi_i} + 
    \sum_{i=4}^{6}  \begin{bmatrix} h_{1i} \\ h_{2i} \end{bmatrix} e^{j \varphi_i} + 
    \sum_{i=7}^{9} \begin{bmatrix} h_{1i} \\ h_{2i}  \end{bmatrix} e^{j \varphi_i} = 
    \begin{bmatrix} 0 \\ 0 \end{bmatrix}.
    \label{eq:spzf-ex1}
\end{equation}
Next, we write the phase rotation as a sum $\varphi_i = \phi_{1i} + \phi_{2j}$, for $i = [1:9]$ and $j = [1:3]$. 
$\phi_{2j}$ corresponds to the $j$-th partition set and is a common factor for terms in the same partition sum.
Thus, we factor $\phi_{2j}$ out of the sums respectively.
\begin{multline}
    e^{j \phi_{21}} \sum_{i = 1}^{3} \begin{bmatrix} h_{1i} \\ h_{2i}\end{bmatrix} e^{j \phi_{1i}} + 
    e^{j \phi_{22}} \sum_{i=4}^{6}  \begin{bmatrix} h_{1i} \\ h_{2i} \end{bmatrix} e^{j \phi_{1i}} \\ +
    e^{j \phi_{23}} \sum_{i=7}^{9} \begin{bmatrix} h_{1i} \\ h_{2i}  \end{bmatrix} e^{j \phi_{1i}} = 
    \begin{bmatrix} 0 \\ 0 \end{bmatrix}.
    \label{eq:spzf-ex2}
\end{multline}
This motivates solving for the $\phi_{1i}$ such that each subset sum of $h_{1i}$ (first row in Equation \ref{eq:spzf-ex2}) is zero. Each partition can be solved independently using the polygon construction method from \cite{zhang_xia_he_zhang_2020}. Let $\{\phi_{1i}^*\}_{i=1}^{N}$ be the phases that zero-force each partition set of $h_1$, then we have
\begin{multline}
    e^{j \phi_{21}}  \begin{bmatrix} 0 \\ \sum_{i = 1}^{3} h_{2i} e^{j \phi_{1i}^*} \end{bmatrix} + 
    e^{j \phi_{22}}  \begin{bmatrix} 0 \\ \sum_{i = 4}^{6} h_{2i} e^{j \phi_{1i}^*} \end{bmatrix} \\ + 
    e^{j \phi_{23}}  \begin{bmatrix} 0 \\ \sum_{i = 7}^{9} h_{2i} e^{j \phi_{1i}^*} \end{bmatrix} = 
    \begin{bmatrix} 0 \\ 0 \end{bmatrix}.
    \label{eq:spzf-ex3}
\end{multline}
When the first row in Equation \ref{eq:spzf-ex3} has been zero-forced and only the second row remains to be solved.
The terms in the second row are obtained by applying the phase rotation $\phi_{1i}^*$ and summing within the partition sets.
Each partition set yields a complex number $y_j$, $j= [1:3]$. 
\begin{equation}
    e^{j \phi_{21}}  \begin{bmatrix} 0 \\ y_1 \end{bmatrix} + 
    e^{j \phi_{22}}  \begin{bmatrix} 0 \\ y_2 \end{bmatrix} + 
    e^{j \phi_{23}}  \begin{bmatrix} 0 \\ y_3 \end{bmatrix} = 
    \begin{bmatrix} 0 \\ 0 \end{bmatrix}.
    \label{eq:spzf-ex4}
\end{equation}
Note, the second row now resembles a single-user zero-forcing problem with channel vector $[y_1, y_2, y_3]$. We can apply the polygon construction method and solve for $\{\phi_{2j}^*\}_{i=1}^{3}$. Setting $\varphi_i = \phi_{1i}^* + \phi_{2j}^*$ for $i = [1:9]$ gives a $\mathbf{w}$ that zero-forces both $\mathbf{h_1}$ and $\mathbf{h_2}$. 

\subsection{Formalized SPZF algorithm for $K = 2$} \label{sse:spzf-k2}
We first formally define the partition and the partition matrix.
\begin{defn}[Partition]
    We denote a partition over the set of $\{h_1, h_2, \dots, h_N\}$ as $\mathcal{B}(\mathbf{h_1}) = \{\mathcal{B}_1, \mathcal{B}_2, \dots, \mathcal{B}_M \}$, which satisfies the following:
    \begin{equation}
        \mathcal{B}_l \subseteq  \{ h_i\}_{i=1}^N,
    \end{equation}
    \begin{equation}
        \bigcup_{l = 1}^{M}  \mathcal{B}_l = \{ h_i \}_{i = 1}^{N},
    \end{equation}
    \begin{equation}
        \mathcal{B}_m \cap \mathcal{B}_n = \emptyset, \forall m \neq n.
    \end{equation}
    We denote the class of all possible partitions as $\mathscr{B}$, i.e. $\mathcal{B}(\mathbf{h_1}) \in \mathscr{B}$. For future purposes, we also denote the subset class of partitions with a fixed cardinality (FC) as $\mathscr{B}_\text{FC}$. Within $\mathscr{B}_\text{FC}$, all partition divides $\mathbf{h_1}$ into sets of exactly $\frac{N}{m}$ elements:
\begin{equation}
    \mathscr{B}_\text{FC} \triangleq \{ \mathcal{B}(\mathbf{h_1}) \; : \; |\mathcal{B}_1| = \dots = |\mathcal{B}_m| = \frac{N}{m} \}.
\end{equation}
    \label{defn:partition}
\end{defn}
\begin{defn}[Partition matrix]
    Let $m$ be the number of partition sets in a specific partition $\mathcal{B}(\mathbf{h_1})$. We denote $\mathbf{B_1} \in \{0,1\}^{m \times N}$ as the partition matrix representing $\mathcal{B}(\mathbf{h_1})$. Let $b_{li}$ be the $li$-th element of $\mathbf{B_1}$:
    \begin{equation}
        b_{li} =
        \left\{
            \begin{array}{ll}
                1, \quad h_{1i} \in \mathcal{B}_l \\
                0, \quad h_{1i} \notin \mathcal{B}_l . \\
            \end{array}
            \right.
        \label{eq:partition-b-defn}
    \end{equation}
\end{defn}
For $K = 2$, the detailed algorithm is described below.
\begin{enumerate}
    \item \emph{Zero-force partitioned $h_1$:} We first partition the terms in $\mathbf{h_1}$ into $m$ sets (construct partition matrix $\mathbf{B_1}$), and apply the polygon construction method to find $\underline{\phi}_1 = [\phi_{11}, \dots, \phi_{1N}]^T$ such that each partition set in the first row of Equation \ref{eq:pzfh1h2} is zero-forced, i.e.
        \begin{equation}
            \mathbf{B_1} \diag(e^{j \underline{\phi}_1}) \mathbf{h_1} = \mathbf{0} \in \mathbb{C}^m.
            \label{eq:pzfh1-partition}
        \end{equation}

    \item \emph{Zero-force $h_2$:} Observe that when Equation \ref{eq:pzfh1-partition} is satisfied, the second row in Equation \ref{eq:pzfh1h2} reduces to:
        \begin{equation}
            \mathbf{y} = \mathbf{B_1} \diag(e^{j \underline{\phi}_1}) \mathbf{h_2},
            \label{eq:reduced-ch-vect}
        \end{equation}
        where $\mathbf{y}$ is a vector in $\mathbb{C}^m$. Provided that the elements in $\mathbf{y}$ satisfy the polygon inequality, we can zero-force $\mathbf{y}$ by finding $\diag(e^{j \underline{\phi}_2})$ such that $\mathbf{y}^T \diag(e^{j \underline{\phi}_2}) = \mathbf{0} \in \mathbb{C}^m$.
        \par Next, we construct the trivial partition matrix $\mathbf{B_2} = [1, \dots, 1] ^T \in \{0,1\}^{1 \times m}$ that maps all elements in $\mathbf{y}$ to one partition. This gives:
        \begin{equation}
            \mathbf{B_2} \diag(e^{j \underline{\phi}_2}) \mathbf{y} = 0.
            \label{eq:pzfh2-partition}
        \end{equation}

    \item \emph{Constructing the final phase:} Combining Equation \ref{eq:reduced-ch-vect} and Equation \ref{eq:pzfh2-partition} and transposing to match $\mathbf{h^T w}$ gives the following solution for Equation \ref{eq:pzfh1h2}: 
        \begin{equation}
            \mathbf{w} = \diag(e^{j \underline{\phi}_1}) \mathbf{B_1}^T \diag(e^{j \underline{\phi}_2}) \mathbf{B_2}^T.
            \label{eq:w-final-form}
        \end{equation}
\end{enumerate}

The pseudocode is presented in Algorithm \ref{alg:successive-pzf-alg}. We denote the polygon construction method from \cite{zhang_xia_he_zhang_2020} as the $\textrm{polygon\_solver}(\cdot)$.
    \begin{algorithm}
        \caption{Successive zero-forcing for $K = 2$}\label{alg:successive-pzf-alg}
        \begin{algorithmic}[1]
            \Require $\mathbf{h_1}$, $\mathbf{h_2}$
            \Require $m$ (number of partition sets)

            \State Partition $\mathbf{h_1}$ and construct partition matrix $\mathbf{B_1}$.
            \For{each set $\mathcal{B}_l$}
            \For{$h_{1i} \in \mathcal{B}_l$}
                $\phi_{1i} \gets  \textrm{polygon\_solver}(\mathcal{B}_l)$
            \EndFor
            \EndFor
            \State $\mathbf{y} \gets \mathbf{B_1} \diag{(e^{j \phi_{1i}})} \mathbf{h_2}$
            \State $\underline{\phi}_2 \gets \textrm{polygon\_solver}(\{y_1, y_2, \dots, y_m\})$
            \State $\mathbf{w} \gets \diag(e^{j \underline{\phi}_1}) \mathbf{B_1}^T \diag(e^{j \underline{\phi}_2}) \mathbf{B_2}^T$
            \Ensure $\mathbf{w}$
        \end{algorithmic}
    \end{algorithm}

The remainder of this paper will focus on partitioning the channel vector to minimize the outage probability for $K = 2$.

\subsection{Theoretical Analysis for the special case of $K = 2$}

\begin{defn}
    \label{defn:reduced-ch-vect}
    Given $\mathbf{h_1}$ and $\mathbf{h_2}$, the reduced channel vector is given by: $\mathbf{y} = \bar{\mathbf{B}} \mathbf{h_2}$, where $\bar{\mathbf{B}} = \mathbf{B_1} \diag{(e^{j \underline{\phi}_1})}$. Without loss of generality, we assume $\bar{\mathbf{B}}$ to be a function of both $\mathbf{h_1}$ and $\mathbf{h_2}$; $\mathbf{B_1}$ is the constructed partition matrix and  $\diag{(e^{j \underline{\phi}_1})}$ is the diagonal matrix with entries equal to the phase rotations that zero-force each partition set of $\mathbf{h_1}$.
\end{defn}





\par By re-arranging the polygon inequality (Equation \ref{eq:pzf-polygon-inequality}), we define a metric to measure how close a partition is to achieving the polygon inequality.
\begin{defn} (\emph{Polygon distance of a set})
    Without loss of generality, assume that $\{|h_i|\}_{i = 1}^N$ is sorted in descending order, i.e. $|h_1| \geq |h_2| \geq \dots \geq |h_N|$, then
        \begin{equation}
            \textrm{dist} (\{|h_i|\}_{i = 1}^N) \triangleq |h_1| - \sum_{i=2}^{N} |h_i|.
            \label{eq:distx}
        \end{equation}
        \label{defn:distx}
    \end{defn}
\begin{remark}
     \label{rmk:dist-interpret}
     When \textrm{dist}($\mathcal{B}$) $\leq 0$, the polygon inequality is achieved.  When \textrm{dist}($\mathcal{B}$) $> 0$, the polygon inequality is not achieved and \textrm{dist}($\mathcal{B}$) is the minimum vector magnitude required to attain the polygon inequality.
\end{remark}

\par We define outage as when the successive zero-forcing approach fails to find a set of phases $\phi_i$ that meet the polygon inequality. This can occur if either error events arise:
    \begin{enumerate}
        \item At least one of the partition sets in channel $\mathbf{h_1}$ fails the polygon inequality.
        \item The reduced channel vector $\mathbf{y}$ fails the polygon inequality.
    \end{enumerate}
Formally, the error events are defined: 
\begin{defn}[Error events]
    \begin{equation}
        \varepsilon_1 \triangleq \{ (\mathbf{h_1}, \mathbf{h_2})  \in \mathbb{C}^{N \times N}: \max_{l = 1, \dots, m} \textrm{dist}(\mathcal{B}_l) > 0 \}.
        \label{eq:err-e1}
    \end{equation}
    Recall that $\mathcal{B}(\mathbf{h_1}) = \{\mathcal{B}_1, \mathcal{B}_2, \dots, \mathcal{B}_m \}$ are the partition sets of $\{h_1\}$.
    \begin{equation}
        \varepsilon_2 \triangleq \{ (\mathbf{h_1}, \mathbf{h_2})  \in \mathbb{C}^{N \times N}: \textrm{dist} (\{|y_l|\}_{l = 1}^m) > 0 \},
        \label{eq:err-e2}
    \end{equation}
    where $y_l$ are the entries of $\mathbf{y}$ according to Definition \ref{defn:reduced-ch-vect}.
    \label{defn:error-prob}
\end{defn}

    In addition, we denote the complement of the two events by $\varepsilon_1^\mathsf{c}$ and $\varepsilon_2^\mathsf{c}$ respectively. The beamforming \emph{outage probability} $\Pr[\mathtt{outage}]$ is the probability that a common beamforming vector cannot be found for the 2-channel system:
    \begin{equation}
        \begin{aligned}
            \Pr [\mathtt{outage}] &= \Pr(\varepsilon_1 \cup \varepsilon_2) \\
            &= \Pr(\varepsilon_1) + \Pr(\varepsilon_2 | \varepsilon_1^\mathsf{c}) \Pr(\varepsilon_1^\mathsf{c}) \\
            & = 1 - \Pr(\varepsilon_1^\mathsf{c}) \Pr(\varepsilon_2^\mathsf{c}| \varepsilon_1^\mathsf{c}).
        \end{aligned}
        \label{eq:total-error-function}
    \end{equation}



Our main research question can be captured succinctly in the following statement:

\begin{prob}
    \label{prob:formal-pzf}
Given two channel vectors $\mathbf{h_1}$, $\mathbf{h_2}$, find a partition $\mathcal{B}(\mathbf{h_1}) = \{\mathcal{B}_1, \mathcal{B}_2, \dots \}$ to minimize $\Pr [\mathtt{outage}]$.
\end{prob}

We first reduce Problem \ref{prob:formal-pzf} by fixing the number of partition sets $m$. We remark that if Problem \ref{prob:formal-pzf-fixed-m} is solved, Problem \ref{prob:formal-pzf} can be solved trivially by iterating through all possible $m$ within the possible $m$ range.
\begin{prob}
    \label{prob:formal-pzf-fixed-m}
    Given two channel vectors $\mathbf{h_1}$, $\mathbf{h_2}$, find a partition with exactly $m$ partition sets $\mathcal{B}(\mathbf{h_1}) = \{\mathcal{B}_1, \mathcal{B}_2, \dots \mathcal{B}_m\}$ to minimize $\Pr [\mathtt{outage}]$.
\end{prob}

\subsection{Problem Transformation for the i.i.d. Rayleigh channel}
In this section, we introduce the following to further reduce the problem: 
\begin{enumerate}
    \item We assume the i.i.d. Rayleigh fading channel. 
    \item We consider the subset $\mathscr{B}_{\text{FC}}$ where each partition set has exactly $\frac{N}{m}$ elements (see Definition \ref{defn:partition}).
\end{enumerate}
We introduce the following notation to denote the outage probability of a channel as a function of the number of channel elements $m$.
\begin{defn}
    The \emph{channel error function} $f(m)$ is the probability distribution for the event that a PZF vector does not exist for a random channel $\mathbf{h}$ with $m$ elements.
    \begin{equation}
        f(m) = \Pr \{ \mathbf{h} \in \mathbb{C}^m: \max_{1 \le i \le m} | h_i| > \frac{1}{2}\sum_{i=1}^{m} |h_i| \}.
    \end{equation}

    \emph{Special case: } If the entries of $\mathbf{h}$ are i.i.d. , $\mathcal{CN}(0,\sigma^2)$, then we denote the channel error function for the Rayleigh i.i.d. channel as $f_\text{Ray} (m)$.  $f_\text{Ray} (m)$ is difficult to evaluate analytically. Lemma \ref{lem:pzf-err-prob} provides a useful asymptotic bound for $f_\text{Ray} (m)$ for large values of $m$.
        \begin{equation}
            f_\text{Ray} (m) \approx m \cdot \exp{(- \frac{\pi}{16} m^2)}.
            \label{eq:fm-error}
        \end{equation}
    \label{defn:channel-error-funct}
\end{defn}
\begin{prop}
    Let $y_l$ be the $l$-th element of $\mathbf{y}$ defined in Definition \ref{defn:reduced-ch-vect}. For a fixed $\bar{\mathbf{B}} = \mathbf{B_1} \diag{(e^{j \underline{\phi}_1})}$ such that all partition sets have the same cardinality $k = \frac{N}{m}$, then $y_l$ are i.i.d. $\mathcal{CN}(0, k\sigma^2)$.
    \label{prop:reduced-ch-vect-iid}
\end{prop}
\emph{Proof.} See Appendix \ref{app:reduced-ch-vect-iid}.
\begin{prop}
    Within $\mathscr{B}_\text{FC}$, $\Pr (\varepsilon_2 | \varepsilon_1^\mathsf{c}) = f_\text{Ray}(m).$
    \label{prop:pr-e2-e1c}
\end{prop}
\emph{Proof.} See Appendix \ref{app:pr-e2-e1c}.
\begin{prop}
    Within $\mathscr{B}_\text{FC}$,
    \begin{equation}
        \arg \min_{\mathcal{B}(\mathbf{h}_1) \in \mathscr{B}_\text{FC}} \Pr[ \mathtt{outage} ] = \arg \min_{\mathcal{B}(\mathbf{h}_1) \in \mathscr{B}_\text{FC}} \Pr(\varepsilon_1).
        \label{eq:minmax}
    \end{equation}
    \label{prop:B-fixed-equivalence}
\end{prop}
\emph{Proof.} From Equation \ref{eq:total-error-function}, we have
    \begin{equation}
    \begin{aligned}
     \Pr[ \mathtt{outage} ] &= 1 - \Pr(\varepsilon_1^\mathsf{c}) \Pr(\varepsilon_2^\mathsf{c} | \varepsilon_1^\mathsf{c})  \\
     &= 1 - \Pr(\varepsilon_1^\mathsf{c}) (1 - f_\text{Ray}(m)).
    \end{aligned}
    \label{eq:outage-B-fixed}
\end{equation}
Note that $f_\text{Ray}(m)$ is fixed for any given $(m, N)$. Thus, over $\mathscr{B}_\text{FC}$, minimizing  $\Pr[ \mathtt{outage} ]$ is equivalent to minimizing $\Pr(\varepsilon_1)$. $\qed$

\par We thus simplify the minimizing objective in Problem \ref{prob:formal-pzf-fixed-m} from $\Pr[ \mathtt{outage} ]$ to $Pr(\varepsilon_1)$. The remainder of the paper will focus on solving Problem \ref{prob:formal-pzf-partition}.

\begin{prob}
    \label{prob:formal-pzf-partition}
    Given channel vector $\mathbf{h_1}$ and the number of partition sets $m$, find a partition $\mathcal{B}(\mathbf{h_1}) = \{\mathcal{B}_1, \mathcal{B}_2, \dots \mathcal{B}_m\}$ that minimizes $\Pr(\varepsilon_1)$.
\end{prob}

Within $\mathscr{B}_{\text{FC}}$,  minimizing $\Pr(\varepsilon_1)$ is equivalent to minimizing $\Pr[\mathtt{outage}]$ (Proposition \ref{prop:B-fixed-equivalence}). Since $\mathscr{B}_{\text{FC}}$ is a subset of $\mathscr{B}$, solving Problem \ref{prob:formal-pzf-partition} minimizes an upper bound  of $\Pr[ \mathtt{outage}]$ over the entire partition class $\mathscr{B}$.
\begin{equation}
    \min_{\mathcal{B}(\mathbf{h_1}) \in \mathscr{B}} \Pr [ \mathtt{outage} ] \leq \min_{\mathcal{B}(\mathbf{h_1}) \in \mathscr{B}_{\text{FC}}} \Pr[ \mathtt{outage} ].
\label{eq:outage-upper-bound}
\end{equation}


\subsection{Objective functions}
In this section, we derive the objective functions that will help drive the optimization algorithm solutions. 

\begin{defn}
    Let $\mathcal{B}(\mathbf{h}) = \{\mathcal{B}_1, \mathcal{B}_2, \dots, \mathcal{B}_m \}$ be a partition of $m$ sets covering all elements of channel $\mathbf{h} = \{ h_i \}_{i = 1}^{N}$, then the loss function $E$ of the partition $\mathcal{B(\mathbf{h})}$ is
    \begin{equation}
        E(\mathcal{B}(\mathbf{h})) = \mathcal{H} [\max_{l = 1, \dots, m} {\textrm dist}(\mathcal{B}_l)],
        \label{eq:loss-function}
    \end{equation}
    where $\mathcal{H}(\cdot)$ is the Heaviside unit step function. 
    \label{defn:loss-function}
\end{defn}
Proposition \ref{prop:loss-to-prob} justifies the choice of this loss function. 
\begin{prop}
    \label{prop:loss-to-prob}
    Let $\Exp{[E]}$ denote the expected loss $E$ over random channel vectors, then $\Pr(\varepsilon_1) = \Exp{[E]}$. Minimizing $\Exp{[E]}$ is equivalent to minimizing $\Pr(\varepsilon_1)$.
\end{prop}
\emph{Proof.} See Appendix \ref{app:loss-to-prob}.

In the following sections we will be presenting iterative algorithms, where the converged loss is of interest. Let $E_k$ denote the loss value at the $k$-th iteration: 
\begin{equation}
    \Pr(\varepsilon_1) = \Exp{[\lim\limits_{k \to \infty} E_k]}.
    \label{eq:iter-part-loss}
\end{equation}
From an algorithmic standpoint, it is more attractive to consider the pseudo-loss by removing the $\mathcal{H}(\cdot)$ from the loss. Proposition \ref{prop:min-e-min-pr} justifies minimizing the pseudo-loss $e$ instead of the true loss $E$. 
\begin{equation}
    e = \max_{l = 1, \dots, m} \textrm{dist}(\mathcal{B}_l).
    \label{eq:pseudo-loss-function}
\end{equation}

\begin{prop}
 Let $e_k$ be the pseudo-loss function at iteration $k$, and $E_k$ be the loss function at iteration $k$.
 If $e_{k+1} \leq e_k$ for all channel vectors, then $\Pr(\varepsilon_1) = \Exp{[\lim\limits_{k \to \infty} E_k]}$ is minimized to a local minimum.
    \label{prop:min-e-min-pr}
\end{prop}
\emph{Proof.} See Appendix \ref{app:min-e-min-pr}.

\section{Proposed Partition Algorithm}
\label{sec:proposed-partition}
In this section, we propose three main algorithms to solve Problem \ref{prob:formal-pzf-partition}. All algorithms aim to minimize $\Pr(\varepsilon_1)$.
For some algorithms, we relax the fixed cardinality constraint and extend the solution space to $\mathscr{B}$ as a heuristic approach. 
Such algorithms are heuristic because it is unknown whether Proposition \ref{prop:B-fixed-equivalence} still holds outside of $\mathscr{B}_\text{FC}$.
The algorithms are:
\begin{enumerate}
    \item Random partition (over $\mathscr{B}_\text{FC})$.
    \item Iterative partition (over $\mathscr{B}$); Iterative partition with a fixed cardinality partition (over $\mathscr{B}_\text{FC}$).
    \item Genetic Algorithm (GA) partition (over $\mathscr{B}$).
\end{enumerate}
\subsection{Random partition} \label{sse:random-partition-alg}
As a baseline, a random partition algorithm is shown below in Algorithm \ref{alg:rand-part}. The primary interest of the random partition is to study the asymptotic behaviour of the outage probability.
    \begin{algorithm}
        \caption{Random partition algorithm}\label{alg:rand-part}
        \begin{algorithmic}[1]
            \Require $\{h_i\}_{i=1}^N$
            \Require $m$ (number of partition sets)
            \State Initialize $m$ empty sets $\mathcal{B}_1, \mathcal{B}_2, \dots, \mathcal{B}_m$.
            \State For all $h_i \in \{h_i\}_{i=1}^N$, randomly assign to one of $\mathcal{B}_1, \mathcal{B}_2, \dots, \mathcal{B}_m$.
            \Ensure $\mathcal{B}_1, \mathcal{B}_2, \dots, \mathcal{B}_m$
        \end{algorithmic}
    \end{algorithm}
A random partition is within $\mathscr{B}_\text{FC}$, hence the outage probability is given by Equation \ref{eq:outage-B-fixed}: $\Pr [\mathtt{outage}] = 1 - \Pr(\varepsilon_1^\mathsf{c})\Pr(\varepsilon_2^\mathsf{c} | \varepsilon_1^\mathsf{c})$.
We further make the following two substitutions. 
First, we note that each partition set $\mathcal{B}_l$ consists of $\frac{N}{m}$ i.i.d. complex numbers drawn from the set $\{h_{1i}\}_{i=1}^{N}$. 
The outage of each $m$ partition sets are jointly independent in a random partition. 
Hence using Definition \ref{defn:channel-error-funct} independently for each partition we have: 
    \begin{equation}
        \Pr(\varepsilon_1^\mathsf{c}) = ( 1- f_\text{Ray}(\frac{N}{m})) ^ m,
        \label{eq:channel-a-part-error}
    \end{equation}
Second, we use Proposition \ref{prop:pr-e2-e1c} directly: 
    \begin{equation}
    \Pr(\varepsilon_2^\mathsf{c}|\varepsilon_1^\mathsf{c}) =  1- f_\text{Ray}(m)
        \label{eq:channel-y-rand}
    \end{equation}
The outage probability is thus:
    \begin{equation}
        \Pr [\mathtt{outage}] = 1 - ( 1- f_\text{Ray}(\frac{N}{m})) ^ m  (1 - f_\text{Ray}(m)).
        \label{eq:random-partition-error}
    \end{equation}

    For a fixed $m$, we show that Equation \ref{eq:random-partition-error} approaches the outage probability of a single channel with $m$ antennas.
    \begin{prop}
        \label{prop:random-part-err-limit}
        Assuming the approximation in Equation \ref{eq:fm-error} in exact, then for a fixed $m$, 
            \begin{equation}
                \lim _{N \to \infty} \Pr [\mathtt{outage}] = f_\text{Ray}(m).
                \label{eq:rand-part-error-limit}
            \end{equation}
    \end{prop}
    \emph{Proof.} See Appendix \ref{app:random-part-err-limit}.

    Hence, the random partition can achieve an arbitrarily low outage probability in the limit of large $N$. However, it may not be sufficient for limited $N$ in practical systems. We propose the following algorithm to partition $\mathbf{h_1}$ more intelligently.


\subsection{Iterative Partition Algorithm} \label{sse:iter-partition-alg}
We propose an iterative algorithm to minimize the pseudo-loss function $e$. The main idea is to move elements from partition sets with negative distances to partition sets with positive distances. We use an iterative approach to minimize the pseudo-loss function to a local minimum. The key steps of the algorithm are as follows:
\begin{enumerate}
    \item Initialization: the algorithm initializes $m$ empty partition sets $\mathcal{B}_1, \dots, \mathcal{B}_m$ and randomly assigns the N elements $h_1, \dots, h_N$ to one of the partition sets $\mathcal{B}_l$.
    \item An \emph{epoch} is defined to be one pass of all partition sets with negative distances. The algorithm minimizes the pseudo-loss function  (Equation \ref{eq:pseudo-loss-function}) by sequentially considering each failing partition ($\mathcal{B}_l$). For a failing partition with distance $\textrm{dist}(\mathcal{B}_l) = d > 0$, we need to add a vector with magnitude greater than or equal to $d$ (Remark \ref{rmk:dist-interpret}). In the ideal case, there is a candidate $h_i$ such that $|h_i|>d$ and $\textrm{dist}(\mathcal{B}_s \setminus \{h_i\}) < 0$; the algorithm moves $h_i$ from $\mathcal{B}_s$ to $\mathcal{B}_l$. This means that we can take a vector $h_i$ from a source partition $\mathcal{B}_s$ without breaking the polygon inequality of $\mathcal{B}_s$. However, if there is no such $h_i$, we find the next-best option and find $h_i^*$ such that
    \begin{equation}
        \begin{aligned}
            & h_i^* = \arg \min_{|h_i| > d} \textrm{dist}(\mathcal{B}_s \setminus h_i), \\
            \text{subject to} \quad & \textrm{dist}(\mathcal{B}_s \setminus h_i) <  d = \textrm{dist}(\mathcal{B}_l).
        \end{aligned}
    \end{equation}
    In the algorithm, these constraints are implemented by pocket values: \texttt{pocket\_d}, \texttt{pocket\_B}, \texttt{pocket\_hi}. \texttt{pocket\_d} is initialized to $\textrm{dist}(\mathcal{B}_l)$ to ensure $\textrm{dist}(\mathcal{B}_s \setminus h_i) <  \textrm{dist}(\mathcal{B}_l)$.
\end{enumerate}

    \begin{algorithm}
        \caption{Iterative partition algorithm}\label{alg:iter-part}
        \begin{algorithmic}[1]
            \Require $\{h_i\}_{i=1}^N$
            \Require $m$ (number of partition sets)

            \State \textbf{Initialize:}
            \Indent
            \State m partition sets $\mathcal{B}_1, \mathcal{B}_2, \dots, \mathcal{B}_m$.
            \State For all $h_i$, randomly assign to one of $\mathcal{B}_1, \mathcal{B}_2, \dots, \mathcal{B}_m$.
            \EndIndent
            \While{ $E(\mathcal{B}_1, \mathcal{B}_2, \dots, \mathcal{B}_m) > 0$ or $E(\mathcal{B}_1, \mathcal{B}_2, \dots, \mathcal{B}_m)$ converges}
            \ForAll{$\mathcal{B}_l$ with $\textrm{dist}(\mathcal{B}_l) > 0$, in increasing order}
            \State d = $\textrm{dist}(\mathcal{B}_l)$
                    \State \texttt{pocket\_d} $\gets d$

                    \ForAll {$\{h_i\}$, $|h_i| > d,$ in increasing order}
                    \State Let $h_i \in \mathcal{B}_s \neq \mathcal{B}_l$
                    \If {$\textrm{dist}(\mathcal{B}_s \setminus \{h_i\})< $ \texttt{pocket\_d} }
                    \State \texttt{pocket\_d} $\gets \textrm{dist}(\mathcal{B}_s \setminus \{h_i\})$
                    \State \texttt{pocket\_B} $\gets \mathcal{B}_s$
                        \State \texttt{pocket\_hi} $\gets h_i$

                        \If {  \texttt{pocket\_d} $ < 0$ }
                        \State \textbf{break for}
                        \EndIf
                    \EndIf
                    \EndFor

                \If {\texttt{pocket\_d} $ < d$}
                    \State $\mathcal{B}_s \gets $ \texttt{pocket\_B}
                    \State $\mathcal{B}_s \gets \mathcal{B}_s \setminus \{\texttt{pocket\_hi}\}$
                    \State $\mathcal{B}_l \gets \mathcal{B}_l \cup \{\texttt{pocket\_hi}\}$
                \EndIf

                \EndFor
            \EndWhile

            \Ensure $\mathcal{B}_1, \mathcal{B}_2, \dots, \mathcal{B}_m$
        \end{algorithmic}
    \end{algorithm}

    \begin{prop}
        \label{prop:loss-monotonic-non-incr}
        The pseudo-loss function $e(\mathcal{B}_1, \mathcal{B}_2, \dots, \mathcal{B}_m)$ is monotonically non-increasing.
    \end{prop}

    \emph{Proof.} At \emph{epoch} k, suppose we have a partition set $\mathcal{B}_1, \mathcal{B}_2, \dots, \mathcal{B}_m$. Consider a partition $\mathcal{B}_l$ with distance $d >0$. The iterative algorithm only updates $\mathcal{B}_l$ if there is a $h_i \in \mathcal{B}_s$, $|h_i| > d$ such that the source partition $\textrm{dist} (\mathcal{B}_s \setminus h_i) < d$, or $\mathcal{B}_l$ stays the same. Suppose $e_k = \textrm{dist} (\mathcal{B}_l)$, then for the first case, we know $e_{k+1} < d = e_{k}$. For the second case, $e_{k+1} = d = e_{k}$. Hence, $e_{k+1} \leq e_{k} \; \forall k$.  $\qed$

    Therefore, by Proposition \ref{prop:min-e-min-pr}, the iterative algorithm minimizes $\Pr(\varepsilon_1)$ locally. The following proposition shows that the iterative algorithm achieves a lower $\Pr(\varepsilon_1)$ than the random partition algorithm.

    \begin{prop}
        \label{prop:better-than-random-err}
        We denote $\Pr [ \varepsilon_1;I ]$ and $\Pr [ \varepsilon_1;R ]$ as the probabilities when the iterative partition algorithm and the random partition algorithm are used, respectively.
        \begin{equation}
            \Pr [\varepsilon_1; I] \leq \Pr [\varepsilon_1; R].
            \label{eq:rand-err-upper-bound}
        \end{equation}
    \end{prop}

    \emph{Proof.} Since the iterative algorithm starts with a random partition, $\Exp{[E_0]}$ (the expected loss value at epoch $0$) is equivalent to the expected final loss value of the random partition algorithm. In addition, from Proposition \ref{prop:min-e-min-pr}, we know that $\Exp[E_{k+1}] \leq \Exp{[E_k]}$ for any $k$. Hence,
    \begin{equation}
        \Pr [\varepsilon_1; I]  = \Exp[\lim\limits_{k \to \infty} E_k] \leq \Exp{[E_0]} = \Pr [\varepsilon_1; R],
    \end{equation}
    which completes the proof. $\qed$

    If we limit the iterative algorithm to only produce partitions within $\mathscr{B}_\text{FC}$, then minimizing $\Pr(\varepsilon_1)$ is equivalent to minimizing the outage probability (Proposition \ref{prop:B-fixed-equivalence}). Let $ \Pr[\mathtt{outage};I_\text{FC}]$ denote the outage probability when the iterative partition with FC is used, then we have: 
        \begin{equation}
            \Pr[\mathtt{outage};I_\text{FC}] \leq \Pr[\mathtt{outage};R].
            \label{eq:outage-IFC-leq-R}
        \end{equation}
        By Proposition \ref{prop:random-part-err-limit}, $\Pr[\mathtt{outage};R]$ approaches $0$ in the limit of large $N$. Hence, the iterative algorithm can also achieve an arbitrarily low outage probability with a smaller number of antennas. We present the iterative algorithm with a fixed cardinality constraint in Appendix \ref{app:iterative-fc-pseudocode}.
    Finally, we present the complexity analysis for the iterative partition.
    \subsubsection{Time Complexity}
    Let $K$ denote the maximum number of epochs required for convergence. In each epoch, the algorithm involves sorting both the $N$ elements and $m$ partition sets, adding a complexity of $\mathcal{O}(N \log N)$ for each epoch. For each of the $m$ partition sets, the algorithm iterates through all elements $|h_i| \geq d$. Since $m$ is bounded by $N$, this contributes an additional $\mathcal{O}(N^2)$ per epoch. The overall time complexity is $\mathcal{O}(K(N \log N + N^2))$.
    \subsubsection{Space Complexity}
        The storage is only needed for 1) the input $\{ |h_i| \}$, and 2) tracking the partition index for each element; both are $\mathcal{O}(N)$. The pocket values contribute to a constant overhead. The overall space complexity is $\mathcal{O}(N)$.

    \subsection{Genetic Algorithm} \label{sse:ga-partition-alg}
   We employ a genetic algorithm (GA) \cite{Memetic} and attempt to find a global minimum of the pseudo-loss function (\ref{eq:pseudo-loss-function}).  The GA is a well-known heuristic approach for addressing non-convex NP-hard problems that cannot be solved in polynomial time. The primary steps of the GA's implementation are summarized below.
    \subsubsection{Overview}
    The GA, inspired by the principles of natural evolution, begins with the initialization of a population, where each individual represents a potential solution to the optimization problem. The core of the algorithm lies in generating new offspring solutions through genetic operators in each generation. These operators, primarily mutation and crossover, introduce variations that simulate the evolutionary process. The minimization of the pseudo-loss value $e$ guides the preservation of a subset of the population for breeding the next generation, similar to natural selection. Unlike the iterative approach that may converge to a local minimum, the GA leverages random permutations in its evolutionary process, enhancing the likelihood of finding a global optimum. The algorithm iteratively evolves the population, terminating either when it achieves a satisfactory pseudo-loss level that remains constant over several generations or when it reaches a predefined maximum number of generations.
    \begin{figure}[htbp]
    \centering
    \begin{tikzpicture}[auto, every node/.style={rectangle, draw, fill=white}]
        \node (node1) {Initialize population};
        \node (node2) [below=0.4cm of node1] {Evaluate fitness, pick elites};
        \node (node3) [below=0.4cm of node2] {Binary tournaments};
        \node (node4) [below=0.4cm of node3] {Select parents};
        \node (node5) [below=0.4cm of node4] {Apply genetic operators}; 
        \node (node6) [below=0.8cm of node5] {Evaluate fitness of offspring};
        \node (node7) [below=0.4cm of node6] {Select individuals for next generation}; 
        \node (node8) [below=0.8cm of node7] {Terminate or repeat};
        \draw [->] (node1) -- (node2);
        \draw [->] (node2) -- (node3);
        \draw [->] (node3) -- (node4);
        \draw [->] (node4) -- (node5);
        \draw [->] (node5) -- node[anchor=west] {\tiny Crossover and mutation} (node6);
        \draw [->] (node6) -- (node7);
        \draw [->] (node7) -- node[anchor=west] {\tiny Check termination conditions} (node8);
        \draw [->] (node8.west) -- ++(-1.8,0) |- (node2.west);
    \end{tikzpicture}
    \caption{Flowchart of the genetic algorithm.}
    \label{fig:tikz}
\end{figure}
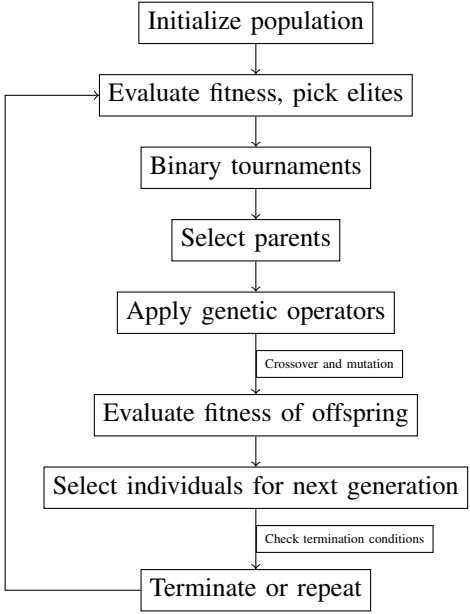

    \subsubsection{Representation}
    The genetic representation of the solution domain will be $N$-dimensional vectors $x \in \{1, \cdots, m\}^N$, where $N$ denotes the length of the channel vector $\mathbf{h} = \{ h_i \}_{i = 1}^{N}$, and $m$ denotes the number of partition sets. These are referred to as chromosomes.
    \begin{figure}[ht]
    \centering
    \begin{tikzpicture}[scale=0.75] 
        \newlength{\nodesize}
        \setlength{\nodesize}{0.75cm} 
        \foreach \x/\value in {1/3,2/1,3/4,4/2,5/3,6/4} {
            \node [draw, minimum width=\nodesize, minimum height=\nodesize] at (\x,0) (node\x) {\value};
            \ifnum\x=1
                \node [below] at (node\x.south) {$\mathcal{B}_2$};
            \else
                \ifnum\x=2
                    \node [below] at (node\x.south) {$\mathcal{B}_1$};
                \else
                    \ifnum\x=3
                        \node [below] at (node\x.south) {$\mathcal{B}_m$};
                    \else
                        \ifnum\x=4
                            \node [below] at (node\x.south) {$\cdots$};
                        \else
                            \ifnum\x=5
                                \node [below] at (node\x.south) {$\mathcal{B}_3$};
                            \else
                                \ifnum\x=6
                                    \node [below] at (node\x.south) {$\mathcal{B}_m$};
                                \fi
                            \fi
                        \fi
                    \fi
                \fi
            \fi
            \ifnum\x=4
                \node [above] at (node\x.north) {$\cdots$};
            \else
                \ifnum\x=5
                    \node [above] at (node\x.north) {$h_{N-1}$};
                \else
                    \ifnum\x=6
                        \node [above] at (node\x.north) {$h_N$};
                    \else
                        \node [above] at (node\x.north) {$h_{\x}$};
                    \fi
                \fi
            \fi
        }

        \node [left] at (node1.west) {Genes};

        \node [above left] at (node1.north west) {Channel Elements};

        \node [below left] at (node1.south west) {Partition Sets};
        
        \node [draw, minimum width=\nodesize, minimum height=\nodesize, fill=white!30] at (1,0) {\textcolor{black}{2}};
        \node [draw, minimum width=\nodesize, minimum height=\nodesize, fill=white!30] at (3,0) {\textcolor{black}{$m$}};
        \node [draw, minimum width=\nodesize, minimum height=\nodesize, fill=white!30] at (4,0) {\textcolor{black}{$\dots$}};
        \node [draw, minimum width=\nodesize, minimum height=\nodesize, fill=white!30] at (6,0) {\textcolor{black}{$m$}};
    \end{tikzpicture}
    \caption{Example of chromosome of a solution. We encode each potential partition solution as a $N$-vector. The $i$-th element takes value from $\{1, \cdots, m\}$, which denotes the partition index that $h_i$ belongs to. } 
    \label{fig:vector}
\end{figure}

    \subsubsection{Minimizing objective}
    We use the same objective as the iterative algorithm and minimize the pseudo-loss function $e = \max_{l = 1, \dots, m} \textrm{dist}(\mathcal{B}_l)$ (Equation \ref{eq:pseudo-loss-function}). By Proposition \ref{prop:min-e-min-pr}, if the pseudo-loss function is monotonically non-increasing, then the algorithm $\Pr(\varepsilon_1)$ is minimized to a local minimum, and so is the upper bound established in Equation \ref{eq:outage-upper-bound}.
    \subsubsection{Initial Population}
    Random generation is widely used for constructing the initial population of chromosomes \cite{Memetic}. Elements of a chromosome $x_i$ can take on any value in $[1, m]$ with equal probability. The population size $P$ is set to $10N$ where $N$ is the length of $\mathbf{h}$.
    \subsubsection{Selection}
    In each generation of the GA, parent solutions are chosen for breeding and mutation based on their loss value, thus enhancing the likelihood of optimal solution convergence. 
    The binary tournament method is used to select parents for breeding \cite{Memetic}. 
    In this method, two candidate solutions are randomly selected. The loss value is compared and the candidate with a lower loss is chosen as the parent for a reproductive trial.
    Each tournament produces one parent. 
    To produce an offspring, two binary tournaments are held. 
    We set an elite count parameter of $25$ to ensure that the $25$ individuals with the lowest pseudo-loss function values are carried over unchanged to the next generation. 
    This ensures a monotonic non-increasing trend in the loss function.
    \subsubsection{Genetic Operators}
    Genetic operators introduce diversity by first creating offspring from existing solutions using crossover, followed by introducing variations through mutation \cite{genetic_operators}. The key distinction is that crossover combines two chromosomes to produce a new offspring, whereas mutation introduces changes within a single offspring. Note that it is not guaranteed that an offspring has a lower loss than its parents.

    \textit{Crossover:} In our experiments, we employed a single-point crossover technique, where a random integer between $1$ and $N-1$ is chosen as the crossover point $k$. Given two parents $p_1$ and $p_2$, offsprings $o_1$ and $o_2$ will be generated where
    \begin{equation}
        \begin{aligned}
            &o_1 = \{p_1[1:k], p_2[k+1:N]\}, \\
            &o_2 = \{p_2[1:k], p_1[k+1:N]\}.
        \end{aligned}
    \end{equation}
    Through our experimental results, we found that an optimal crossover rate is $85\%$. This implies that $85\%$ of the new generation is created through crossover while the remaining $15\%$ are direct copies from the previous generation. There is a trade-off between a high crossover rate (faster convergence) and population diversity.

    \textit{Mutation:} Mutation modifies one or more elements in each new offspring, which helps prevent the population from converging to the local minimum. The mutation operator moves the genes (elements) of a chromosome from partition $i$ to partition $j$, where $i, j \in \{1, ..., m\}$. We set the mutation to occur with a 10\% probability at each gene in a chromosome.
    \begin{algorithm}
        \caption{Genetic Algorithm}\label{alg:GA}
        \begin{algorithmic}[1]
            \Require $\{h_i\}_{i=1}^N$
            \Require {Population size, $P$}. Maximum number of iterations, $K$

            \State \textbf{Initialize:}
            \Indent
            \State Generate initial population of $n$ chromosomes $x_{i}, i = 1, 2, \ldots, n$
            \State Set iteration counter $k$ to $0$
            \EndIndent
            \While{$k < K$}
            \State Evaluate the loss value of each chromosome
            \State Select a pair of chromosomes based on loss (binary tournament)
            \State Perform crossover operation on selected chromosomes based on crossover probability
            \State Perform mutation on the offspring based on mutation probability
            \State Replace the old population with the newly generated population
            \State Increment the iteration $k$ by $1$
            \EndWhile
            \Ensure $x_\text{min}$, candidate solution with lowest pseudo-loss $e$.
        \end{algorithmic}
    \end{algorithm}
    The complexity analysis of the GA is as follows. 
    \setcounter{subsubsection}{0}
    \subsubsection{Time Complexity} \hfill\\
        \textit{Initialization:} The initial population is randomly generated, assuming each individual is represented by an $N$-dimensional vector and is generated in constant time. If the population size is $P$, this operation has complexity $\mathcal{O}(PN)$.

        \textit{Selection:}
        The selection process requires the evaluation of the loss function of every individual in a population, where each individual is evaluated in constant time. Therefore, this has complexity $\mathcal{O}(PN)$.

        \textit{Crossover:}
        Since we perform single-point crossover, the slicing and concatenation operations are dependent on the length of both individual chromosomes. Each chromosome is an $N$-dimensional vector, where $N$ denotes the length of the channel vector $\mathbf{h}$. The worst-case complexity is $\mathcal{O}(2P N)$.

        \textit{Mutation:}
        Similarly, we perform a mutation on each gene with a certain probability, and therefore the operation is dependent on the length of the chromosome $N$. The worst-case complexity is $\mathcal{O}(P N)$.

        Considering all of the above operations, the time complexity for one generation (selection, crossover, and mutation operations) of the GA is $\mathcal{O}(5PN)$. If the GA runs for $G$ generations, and we omit all constants, then the overall time complexity becomes $\mathcal{O}(G P N)$.
    \subsubsection{Space Complexity}
    The main storage required by the GA is 1) the storage of the population in each generation which is proportional to $\mathcal{O}(P N)$, and 2) the storage of the loss value of each individual which is proportional to $\mathcal{O}(P)$. The genetic operations modify existing chromosomes and do not add to the space complexity. Storing loss values and genetic operator probabilities adds constant overhead. Considering all of the above operations, the space complexity for one generation of the GA is $\mathcal{O}(P N)$. Even if the GA runs for $K$ generations, the new population replaces the old one in memory, and therefore the space complexity remains as $\mathcal{O}(P N)$.

\section{Simulation Results}
\label{sec:simulation-results}
In this section, we validate the theoretical results using Monte-Carlo simulations for the i.i.d. Rayleigh fading channel and the geometric channel with $L = 10$ paths. 
In parts \ref{sec:error-outage-prob} to \ref{sec:beyond-Rayleigh}, we analyze $\Pr(\varepsilon_1)$ and $\Pr[\mathtt{outage}]$ of the proposed SPZF scheme with the random, iterative, and genetic partition algorithms.  
In part \ref{sec:secrecy-rate}, we compare the achievable secrecy rate of the different partition algorithms. All numerical simulations were conducted with $10000$ trials. 
We assume unit variance ($\sigma^2 = 1$) for all channel gains in $\mathbf{h_1}, \mathbf{h_2}$, and $\mathbf{G}$.

\subsection{Asymptotic outage probability}
\label{sec:error-outage-prob}
\begin{figure}[htpb]
    \centering
    \includegraphics[width=\linewidth]{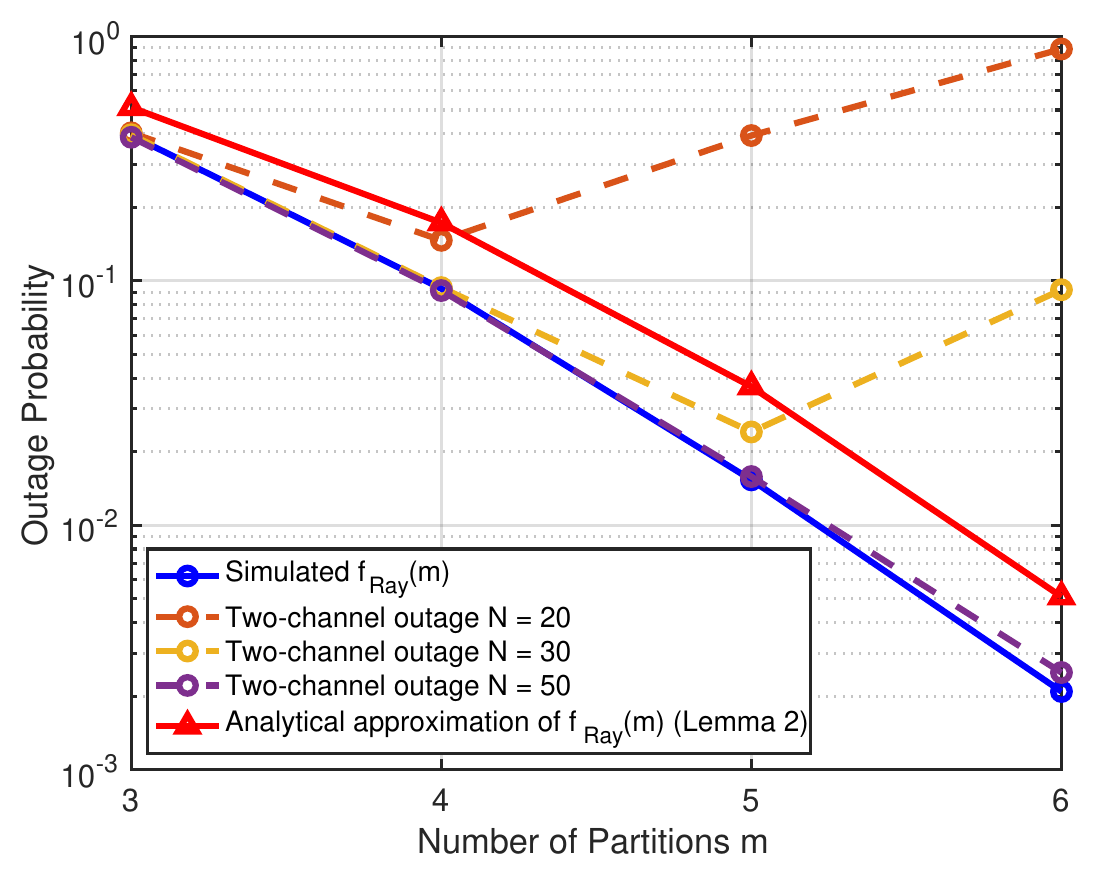}
    \caption{Comparison of $\Pr[\mathtt{outage}]$ of the SPZF using a random partition v.s. $f_\text{Ray}(m)$ outage probability of a Rayleigh i.i.d. fading channel with $m$ terms. The two-channel outage when $N = 50$ matches the empirical $f_\text{Ray}(m)$.}
    \label{fig:random-partition-limit}
\end{figure}

We first demonstrate the outage probability using our proposed SPZF scheme with a random partition. 
In Proposition \ref{prop:random-part-err-limit}, we showed $\lim _{N \to \infty} \Pr [\mathtt{outage}] = f_\text{Ray}(m)$. 
We plot the simulated two-channel $\Pr [\mathtt{outage}]$ as a function of the number of partition sets $m$ in Figure \ref{fig:random-partition-limit}. 
We compare the two-channel $\Pr [\mathtt{outage}]$ for $N = 20, 30, 50$ against the simulated empirical single-channel outage $f_\text{Ray}(m)$ with $m$ antennas.
The results confirm Proposition \ref{prop:random-part-err-limit}. The two-channel $\Pr [\mathtt{outage}]$ approaches $f_\text{Ray}(m)$ for sufficiently large $N$.
We also plot the analytical approximation of $f_\text{Ray}(m)$ from Lemma \ref{lem:pzf-err-prob}, which shows larger values than the empirical $f_\text{Ray}(m)$ for the range $m \in [3:6]$.

\subsection{Comparison of the partition algorithms: $\Pr(\varepsilon_1)$} \label{sec:comp-partition-algs}
\begin{figure}
    \centering
    \subfloat[$\Pr(\varepsilon_1)$, $N = 20$\label{fig:sim-e1-N20}]{%
        \includegraphics[width=0.485 \linewidth]{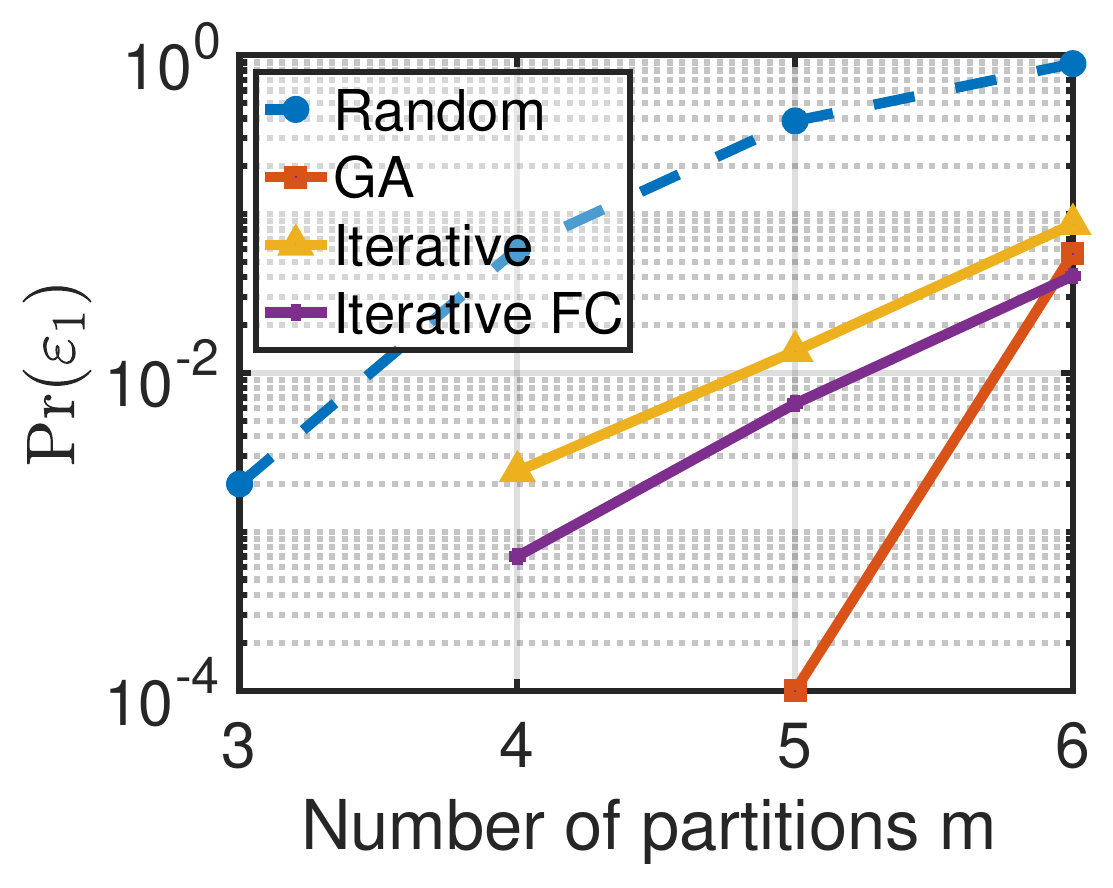}
    }
    \hfill
    \subfloat[ $\Pr{[} \mathtt{outage} {]}$, $N = 20$\label{fig:sim-outage-N20}]{%
        \includegraphics[width=0.485 \linewidth]{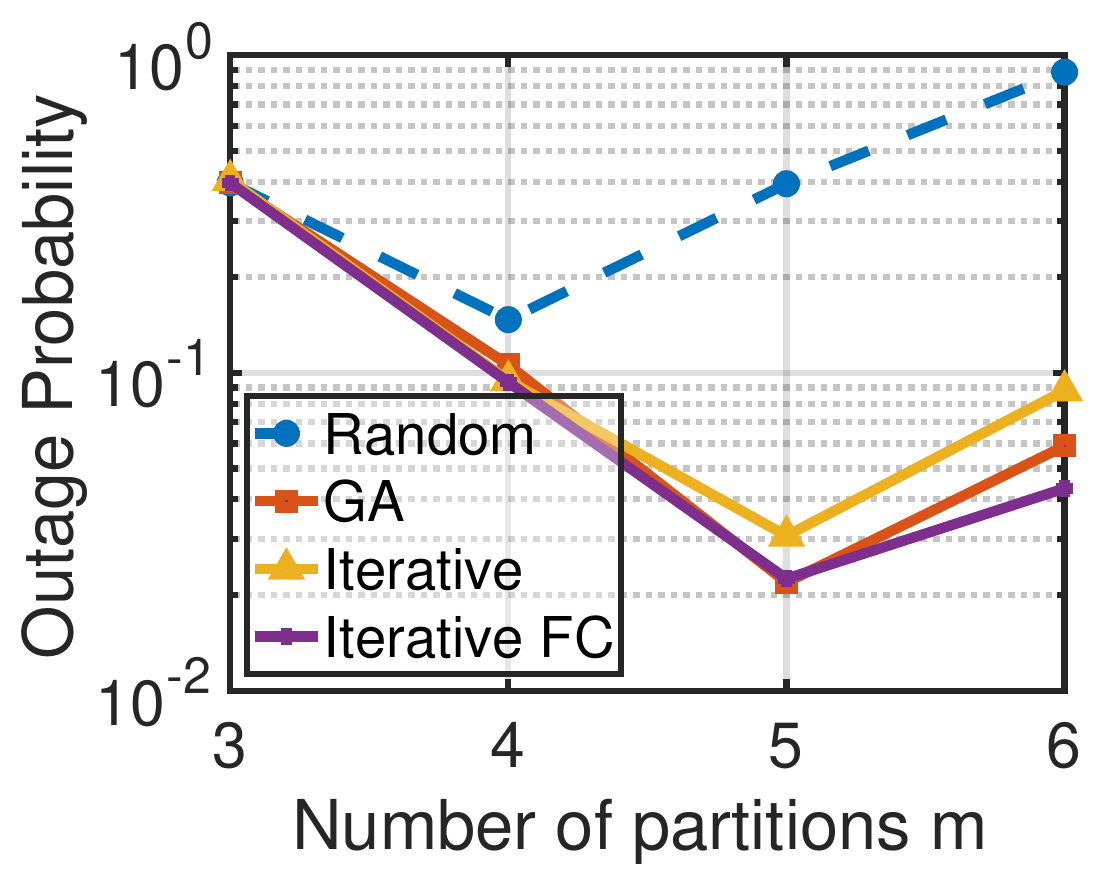}
    }
    \\
    \subfloat[$\Pr(\varepsilon_1)$, $N = 30$\label{fig:sim-e1-N30}]{%
        \includegraphics[width=0.485 \linewidth]{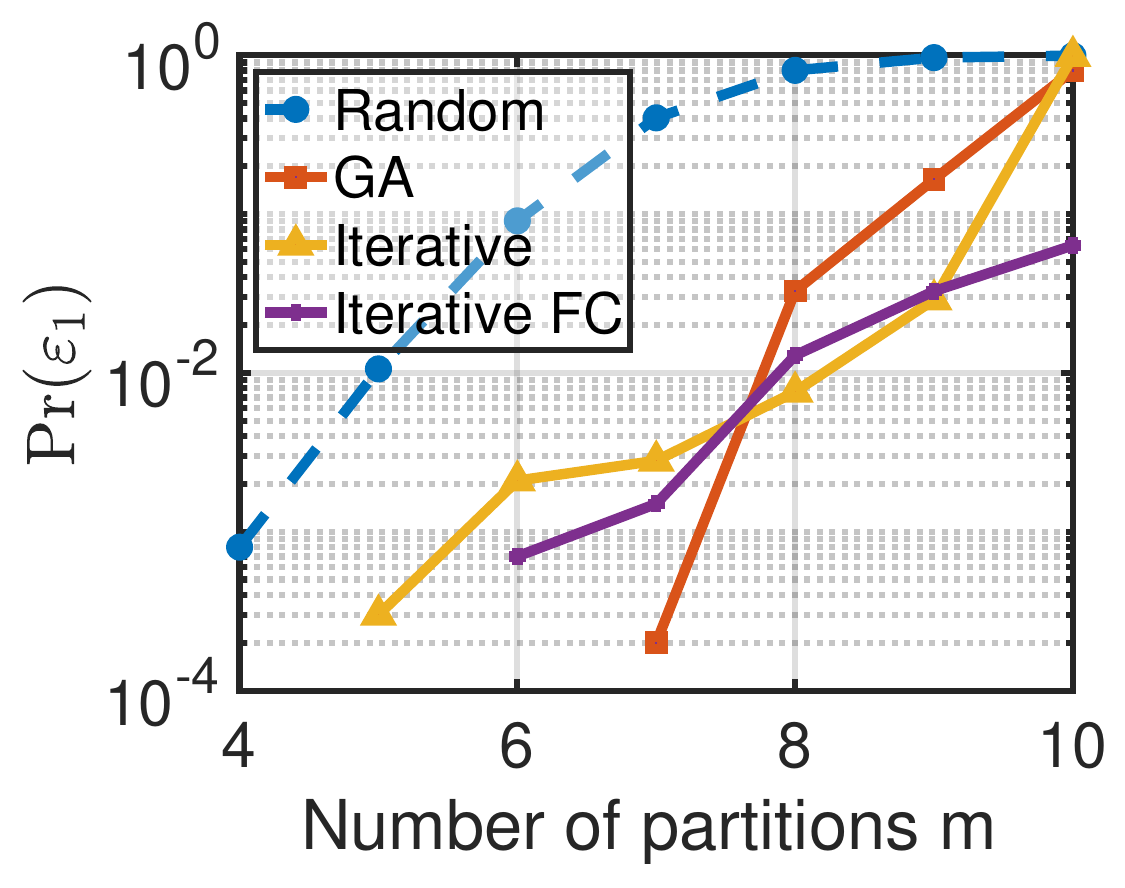}
    }
    \hfill
    \subfloat[ $\Pr {[} \mathtt{outage} {]}$, $N = 30$\label{fig:sim-outage-N30}]{%
        \includegraphics[width=0.485 \linewidth]{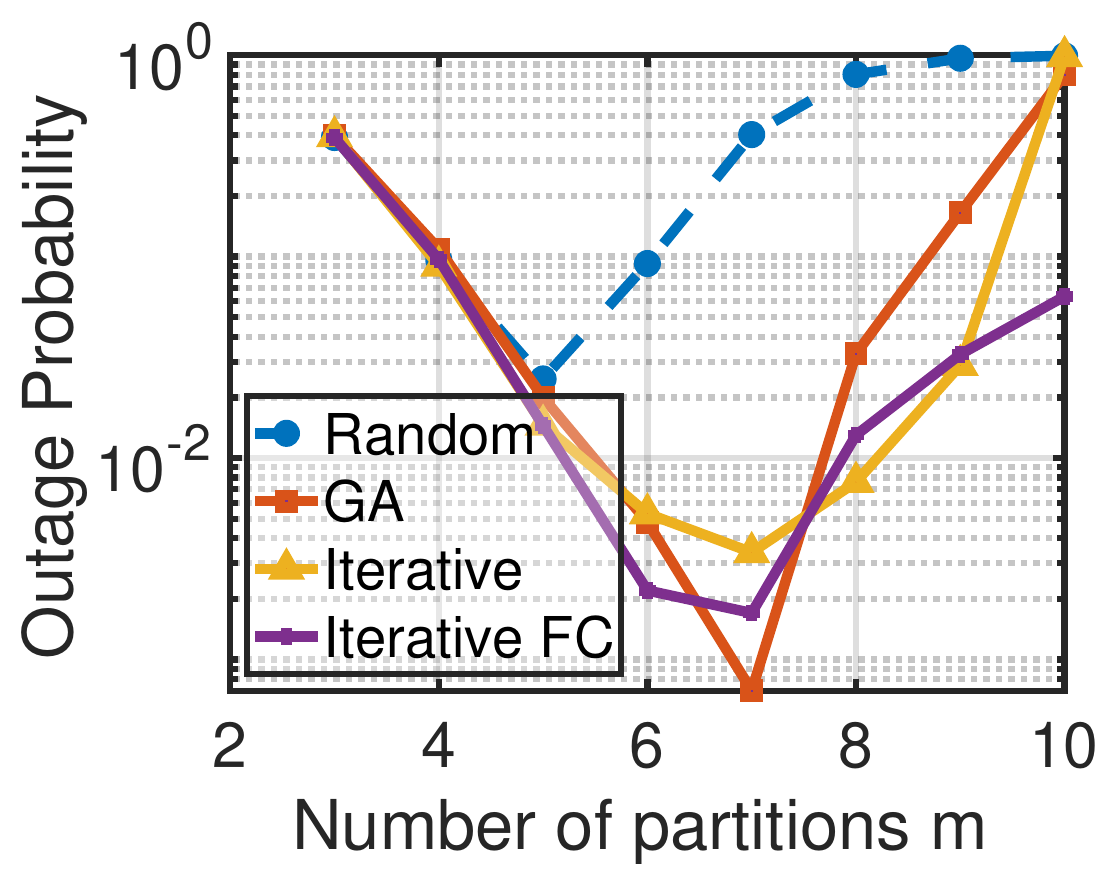}
    }
    \caption{$\Pr(\varepsilon_1)$ and $\Pr[\mathtt{outage}]$ for the i.i.d. Rayleigh fading channel when $N = 20$ and $N = 30$.}
\end{figure}

The three proposed partition algorithms are designed to minimize $\Pr(\varepsilon_1)$.
We plot $\Pr(\varepsilon_1)$ achieved by the three partition algorithms as a function of the number of partition sets $m$ for $N = 20, 30$ in Figure \ref{fig:sim-e1-N20} and Figure \ref{fig:sim-e1-N30}.


Recall that $\Pr(\varepsilon_1)$ indicates the event where any partition set of $\mathbf{h_1}$ fails the polygon inequality.
As $m$ increases, $\Pr(\varepsilon_1)$ increases for all partition algorithms. 
For the random partition algorithm, this can be easily verified by $\Pr(\varepsilon_1) = 1 - (1 - f_\text{Ray}(\frac{N}{m}))^m$ (Equation \ref{eq:channel-a-part-error}), and using the fact that $f_\text{Ray}(m)$ increases as $m$ increases (see Figure \ref{fig:random-partition-limit}). 
For the iterative and genetic algorithms, we do not derive an analytical expression for $\Pr(\varepsilon_1)$, but we have observed the same increasing behaviour as $m$ increases.
For fixed $N$, the average cardinality of each partition set is $\frac{N}{m}$. As $m$ increases, the average cardinality decreases and the polygon inequality becomes harder to satisfy for each of the $m$ sets. Larger $m$ also means more partitions are required to simultaneously satisfy the polygon inequality.


\emph{Comparison of algorithm on minimizing $\Pr(\varepsilon_1)$}: Both the iterative partition (with and without FC) and the GA partition algorithms yield a lower $\Pr(\varepsilon_1)$ than that of the random partition algorithm. 
Notably, we expect the iterative partition algorithm (minimizing over the entire set $\mathscr{B}$) to have a lower $\Pr(\varepsilon_1)$ than the iterative FC partition algorithm (minimizing over the subset $\mathscr{B}_\text{FC} \subset \mathscr{B}$). 
However, in many cases (e.g. Figure \ref{fig:sim-e1-N20}) the iterative FC yields a smaller $\Pr(\varepsilon_1)$ than that of the unconstrained iterative algorithm. 
This might be because both algorithms minimize $\Pr(\varepsilon_1)$ locally. It appears that the unconstrained iterative algorithm does not encounter $\mathscr{B}_\text{FC}$.


\subsection{Comparison of the partition algorithms: $\Pr[\mathtt{outage}]$} \label{sec:comp-partition-outage}
We analyze the outage probabilities achieved by the proposed algorithms and present the results in Figure \ref{fig:sim-outage-N20} for $N = 20$ and Figure \ref{fig:sim-outage-N30} for $N = 30$. 
Contrary to the monotonically increasing behavior of $\Pr(\varepsilon_1)$ with respect to $m$, $\Pr[\mathtt{outage}]$ displays high outage probabilities at both small and large values of $m$. 
This can be explained through Equation \ref{eq:total-error-function}, where $\Pr [\mathtt{outage}] = \Pr(\varepsilon_1) + \Pr(\varepsilon_2 | \varepsilon_1^\mathsf{c}) \Pr(\varepsilon_1^\mathsf{c})$. 
For large values of $m$, $\Pr(\varepsilon_1)$ becomes the dominant term, resulting in the observed high probabilities akin to those seen in the $\Pr(\varepsilon_1)$ plots. 
For small values of $m$, $\Pr(\varepsilon_2 | \varepsilon_1^\mathsf{c})$ dominates and is high because $m$ corresponds to the number of elements in the reduced channel vector $\mathbf{y}$.

We observe a strong correlation between minimizing $\Pr(\varepsilon_1)$ and minimizing $\Pr[\mathtt{outage}]$ across all algorithms. 
This correlation is expected for algorithms with the FC constraint (random, iterative FC), as theoretically shown in Proposition \ref{prop:B-fixed-equivalence}. Notably, we also observe a strong correlation between minimizing $\Pr[\mathtt{outage}]$ and minimizing $\Pr(\varepsilon_1)$ in the heuristic unconstrained algorithms.

\begin{figure}
    \centering
    \subfloat[Rayleigh i.i.d. channel\label{fig:system-min}]{%
        \includegraphics[width=0.485\linewidth]{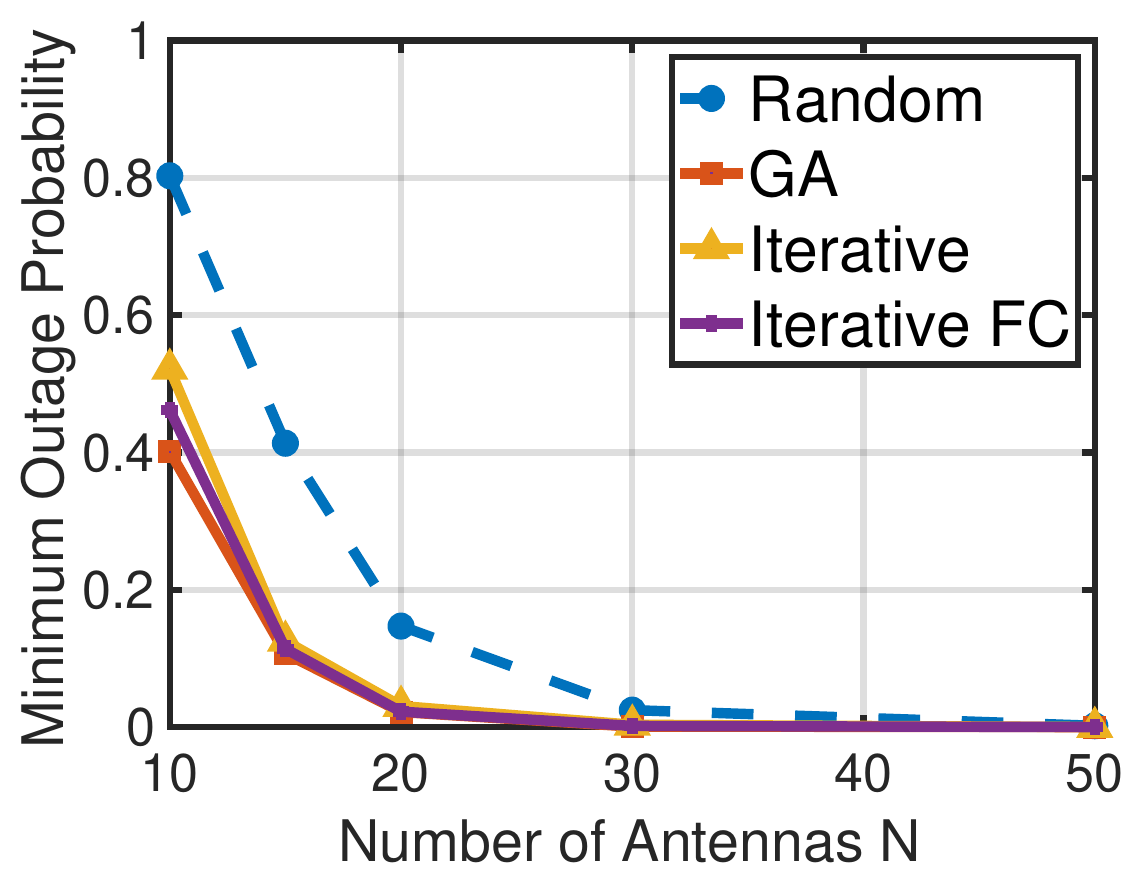}
    }
    \hfill
    \subfloat[Geometric channel with $L = 10$ paths \label{fig:system-min-geometric}]{%
    \includegraphics[width=0.485\linewidth]{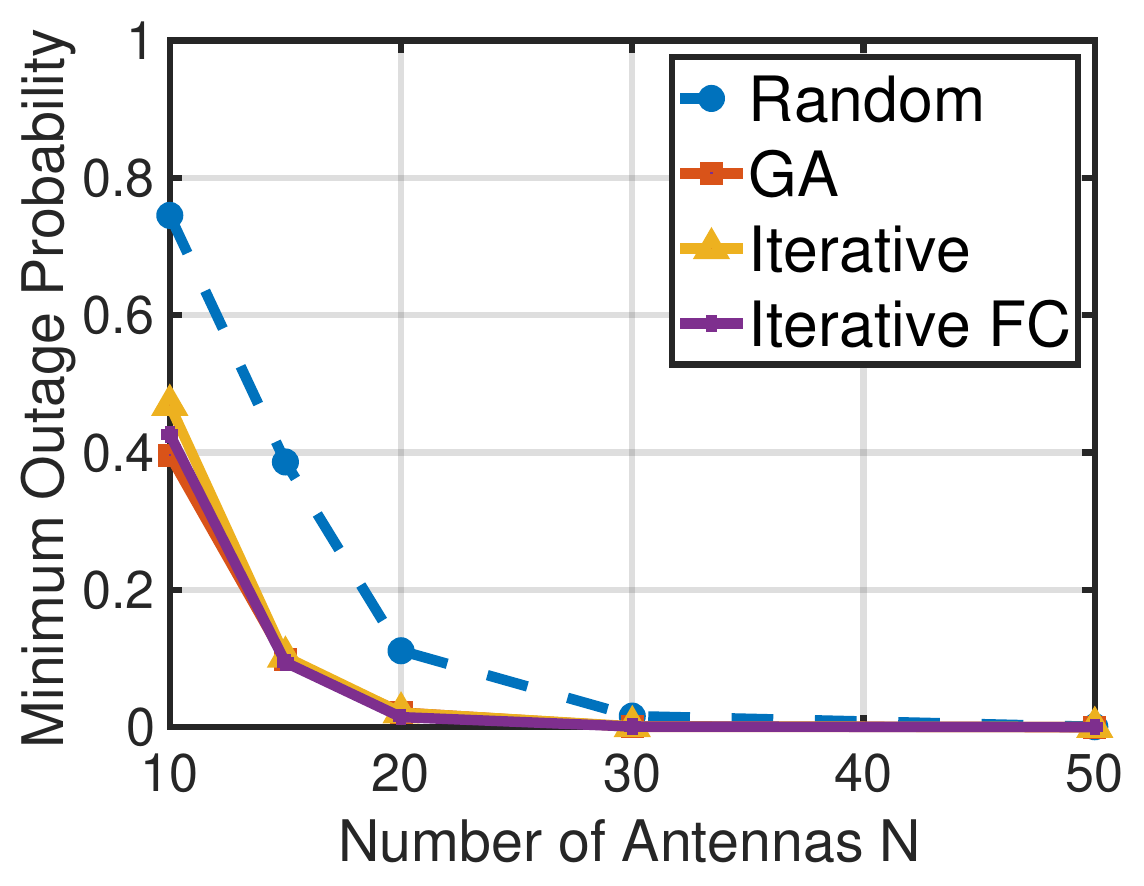}
}
\caption{Achieved minimum $\Pr[\mathtt{outage}]$: the minimum outage probability over $m$ for $N=10,20,30,40,50$. The difference between the baseline random partition and advanced algorithms (iterative, genetic) is most prominent for smaller $N$.}
\end{figure}

Figure \ref{fig:sim-outage-N20} and Figure \ref{fig:sim-outage-N30} suggest that for fixed $N$, there is an optimal number of partition sets $m^* = \arg \min_{m} \Pr[\mathtt{outage}]$. 
For each $N \in \{10, 20, 30, 40, 50\}$, we iterate through $m$ and find the empirical $\min_{m} \Pr[\mathtt{outage}]$.
We plot this simulated minimum outage probability as a function of $N$ in Figure \ref{fig:system-min} for the Rayleigh i.i.d. channel, which illustrates that all partition algorithms approach zero outage probability as $N$ increases in our simulations of 10000 trials. 
The iterative and GA partition algorithms achieve very similar outage probabilities. The improvement in the outage probability of these two algorithms relative to the random partition is most pronounced when $N$ is small.  

\subsection{Beyond i.i.d. Rayleigh Fading} \label{sec:beyond-Rayleigh}
\begin{figure}
    \centering
    \subfloat[$\Pr(\varepsilon_1)$, $N = 20$\label{fig:sim-e1-N20-geometric}]{%
        \includegraphics[width=0.485 \linewidth]{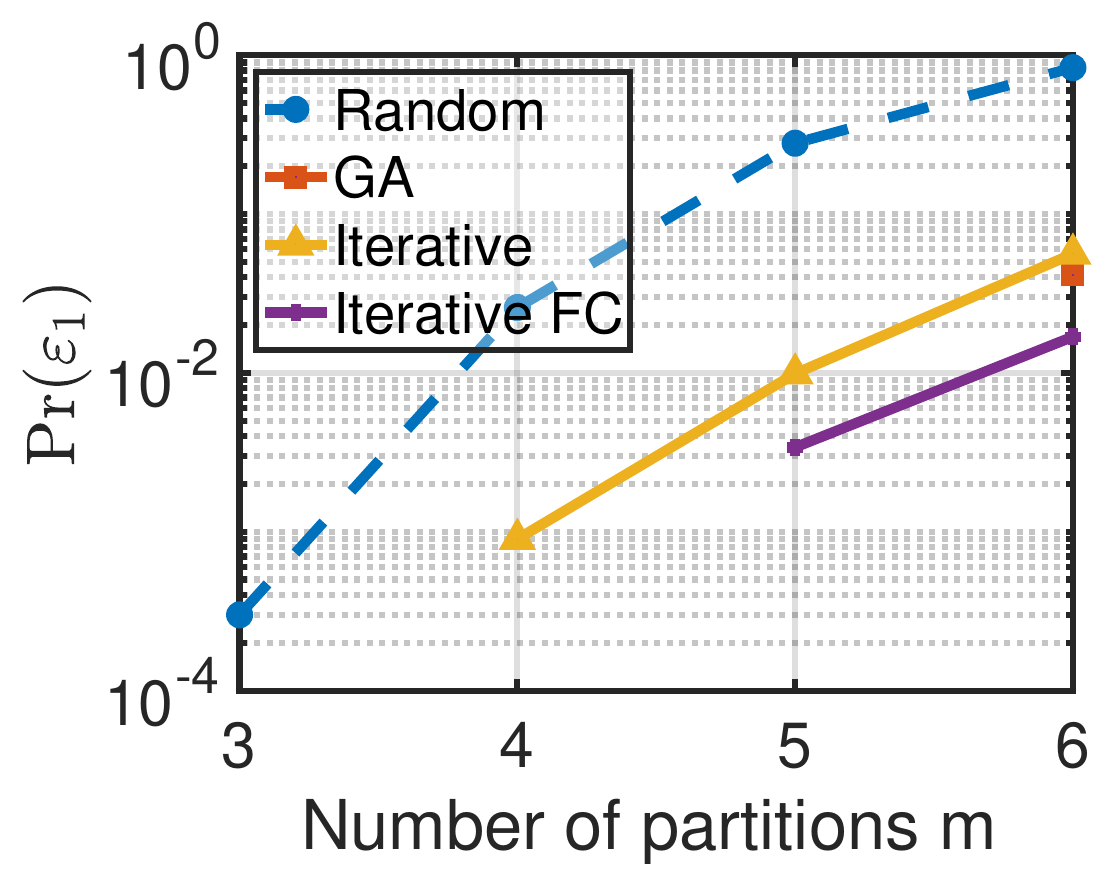}
    }
    \hfill
    \subfloat[$\Pr{[} \mathtt{outage} {]}$, $N = 20$\label{fig:sim-outage-N20-geometric}]{%
        \includegraphics[width=0.485 \linewidth]{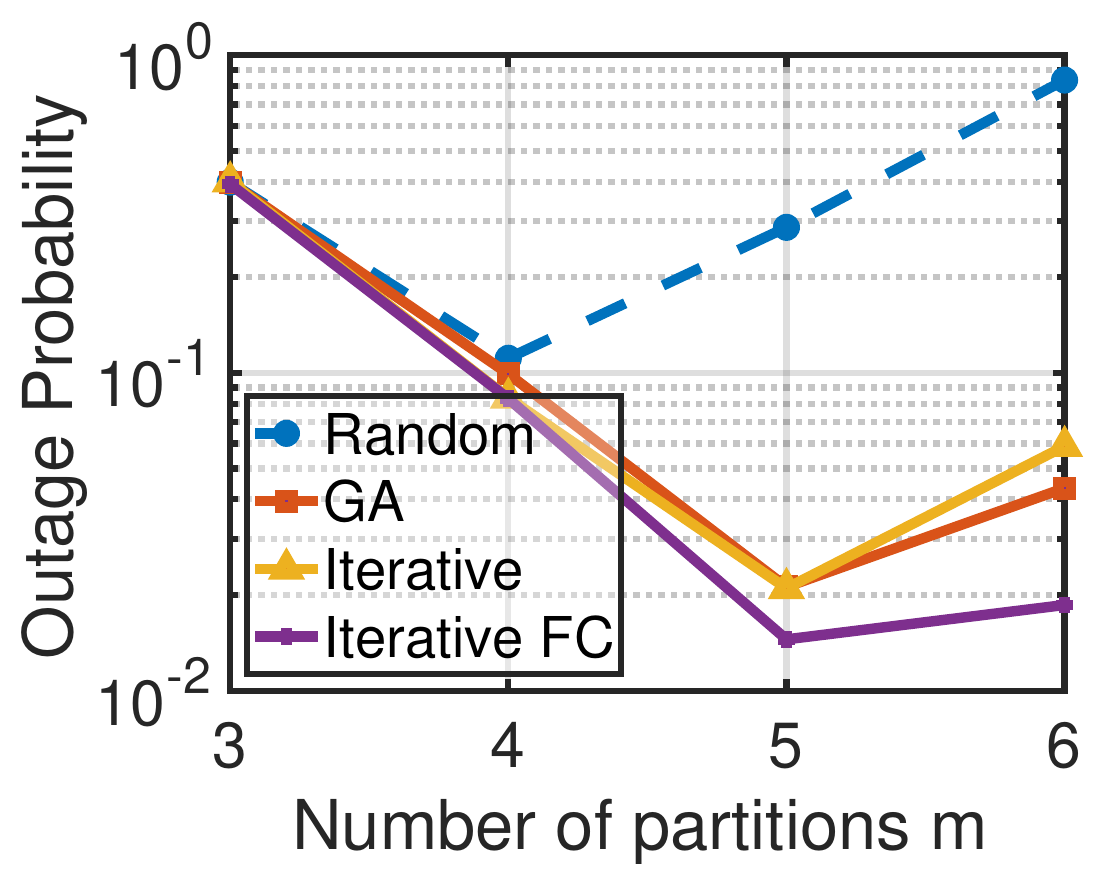}
    }
    \\
    \subfloat[$\Pr(\varepsilon_1)$, $N = 30$\label{fig:sim-e1-N30geometric}]{%
        \includegraphics[width=0.485 \linewidth]{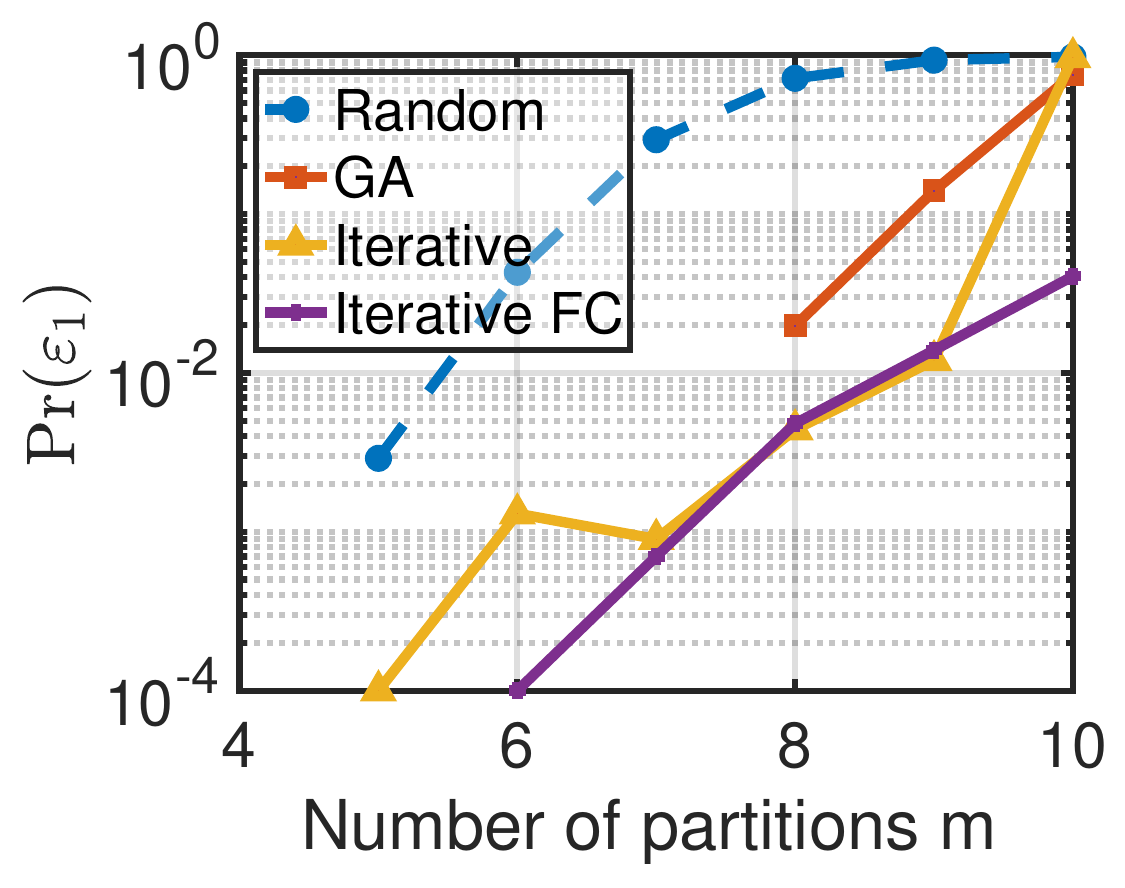}
    }
    \hfill
    \subfloat[$\Pr{[}\mathtt{outage}{]}$, $N = 30$\label{fig:sim-outage-N30-geometric}]{%
        \includegraphics[width=0.485 \linewidth]{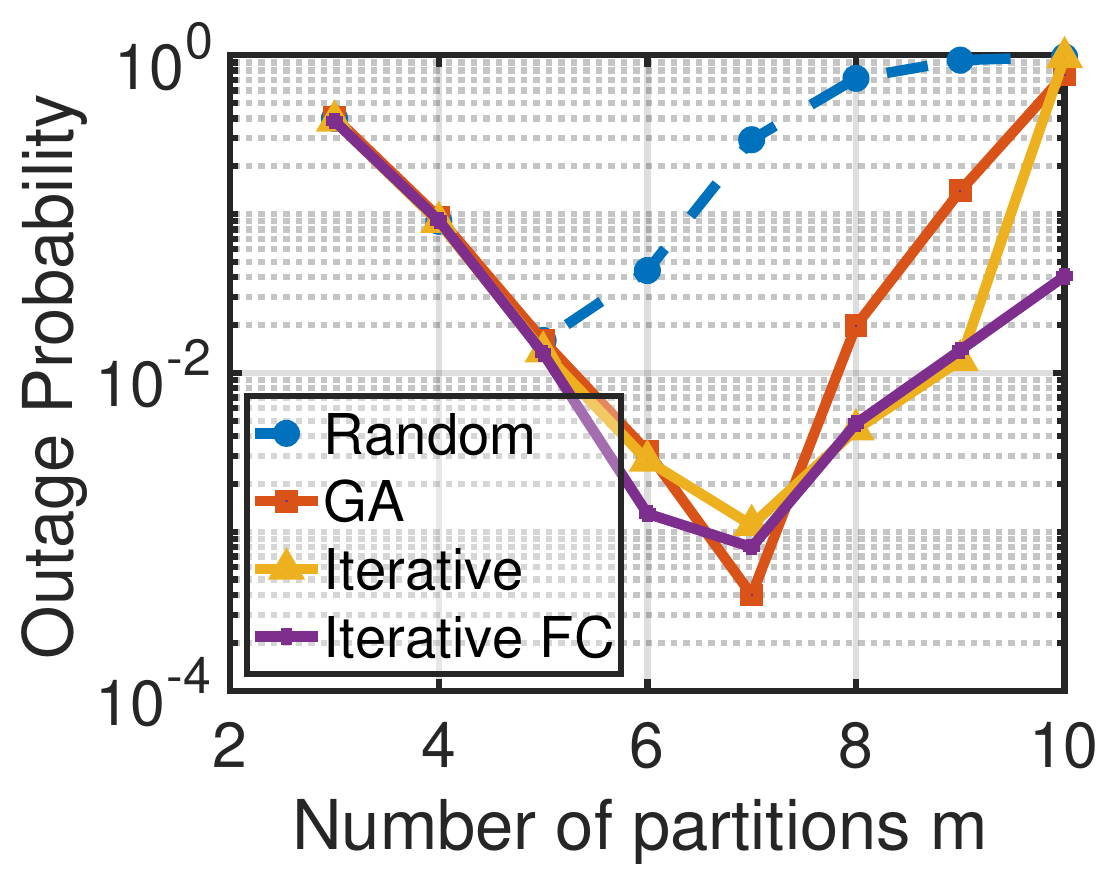}
    }
  \caption{$\Pr(\varepsilon_1)$ and $\Pr[\mathtt{outage}]$ for the geometric channel with $L = 10$ paths when $N = 20$ and $30$.}
  \label{fig:sim-geometric}
\end{figure}

Figure \ref{fig:sim-geometric} illustrates $\Pr(\varepsilon_1)$ and $\Pr[\mathtt{outage}]$ for the geometric channel model with $L = 10$, while Figure \ref{fig:system-min-geometric} displays the minimum outage probability achieved using $m^*$.
For both $\Pr(\varepsilon_1)$ and $\Pr[\mathtt{outage}]$, the geometric channel exhibits trends similar to the Rayleigh i.i.d. channel.
However, a notable difference is observed: for the same parameters $(m, N)$, the geometric channel yields lower values of both $\Pr(\varepsilon_1)$ and $\Pr[\mathtt{outage}]$ than the Rayleigh i.i.d. channel. This observation is consistent with prior research such as \cite{zhao_lee_khisti_2016}, which suggests that the correlation between channel gains in geometric channels can be advantageous for satisfying the polygon inequality.

\subsection{Secrecy Rate Analysis} \label{sec:secrecy-rate}
We next present an analysis of the secrecy rate as a function of the signal-to-noise ratio (SNR) to compare the ability of each algorithm in part IV to find the optimal PZF vector in the null space of the legitimate receiver's channel.  The secrecy rate is defined as the maximum rate at which confidential information can be transmitted through a channel while ensuring that an eavesdropper cannot decode the information with a high probability \cite{secrecy_rate}. 
Mathematically, it is the difference in the capacity of the legitimate receiver channels and the capacity of the eavesdropper's channel \cite{secrecy_rate} (\cite{Csiszar_Korner}, \cite{El_Gamal}): 
\begin{align}
    R &= \min \{ R_1, R_2 \} \nonumber \\
    &= \min \left\{ \begin{array}{l}
            I(\mathbf{x}; y_1 | \mathbf{h_1}) - I(\mathbf{x}; \mathbf{z} | \mathbf{G}), \\
            I(\mathbf{x}; y_2 | \mathbf{h_2}) - I(\mathbf{x}; \mathbf{z} | \mathbf{G})
    \end{array} \right\}, 
\end{align}
where $y_1, y_2, \mathbf{z}, \mathbf{h_1}, \mathbf{h_2}, \mathbf{G}$ are defined in the system model section (c.f. Equation \ref{eq:k-receiver-signal} and \ref{eq:e-eve-signal}), $\mathbf{x}$ includes both the transmit message and the artificial noise, as discussed in the next paragraph.

Our system model involves a single RF chain for transmitting the message $q$ and a separate message for transmitting the constructed artificial noise symbol $s$. Let $q$ have zero mean and unit variance, and let $s \sim \mathcal{CN}(0,1)$.
Let the analog beamforming vector (phase-only) of $q$ and $s$ be $\mathbf{v}$ and $\mathbf{w}$ respectively. 
The phases of each element in $\mathbf{v}$ compensate for the phase of the corresponding legitimate channel gain. 
We construct $\mathbf{w}$ using our proposed beamforming algorithms presented in Sections \ref{sse:random-partition-alg}, \ref{sse:iter-partition-alg}, and \ref{sse:ga-partition-alg}.
The input vector $\mathbf{x}$ then takes the following form
    \begin{equation}\label{eq:10}
        \begin{aligned}
            & \mathbf{x} = \sqrt{\frac{P}{N (M+1)}} \left( q\mathbf{v} + s \mathbf{w} \right),
        \end{aligned}
    \end{equation}
where $M$ is the number of RF chains used for noise transmission (in our case, 1). 

The maximum achievable secrecy rate is derived in \cite[Equation 37]{zhao_lee_khisti_2016}. 
We adapt the expression for using a single RF chain to send a single noise symbol. 
In addition, we re-write the phase-only beamforming vectors $\mathbf{v}$ and $\mathbf{w}$ from Equation \ref{eq:10} to include the power pre-factor, i.e.  $ \mathbf{\tilde{v}} \triangleq \sqrt{\frac{P}{N (M+1)}} \mathbf{v}$ and $\mathbf{\tilde{w}} \triangleq \sqrt{\frac{P}{N (M+1)}} \mathbf{w}$.
Then, the achievable secrecy rate can be expressed as
    \begin{equation}
        \begin{split}
            \label{eq:11}
            & R = \mathbb{E}_{\mathbf{h}, \mathbf{G}} \: \log\left( 1 + \frac{\left| \mathbf{h}^T \mathbf{\tilde{v}}\right|^2}{1 + \left| \mathbf{h}^T \mathbf{\tilde{w}}\right|^2} \right) \\
            & - \log \left( \frac{\det\left( \mathit{I} + \mathbf{G} \mathbf{\tilde{v}} \mathbf{\tilde{v}}^{\dag} \mathbf{G}^{\dag} + \mathbf{G} \mathbf{\tilde{w}} \mathbf{\tilde{w}}^{\dag} \mathbf{G}^{\dag} \right)}{\det\left( \mathit{I} + \mathbf{G} \mathbf{\tilde{w}} \mathbf{\tilde{w}}^{\dag} \mathbf{G}^{\dag} \right)} \right).
        \end{split}
    \end{equation}
We assume the white Gaussian noise to have unit variance, and hence the SNR of the system is $\frac{P}{1} = P$.

\begin{figure}
    \centering
    \subfloat[$N_e = 1$, Rayleigh i.i.d. channel\label{fig:secrecy_N=20_Ne=1-Rayleigh}]{%
        \includegraphics[width=0.485\linewidth]{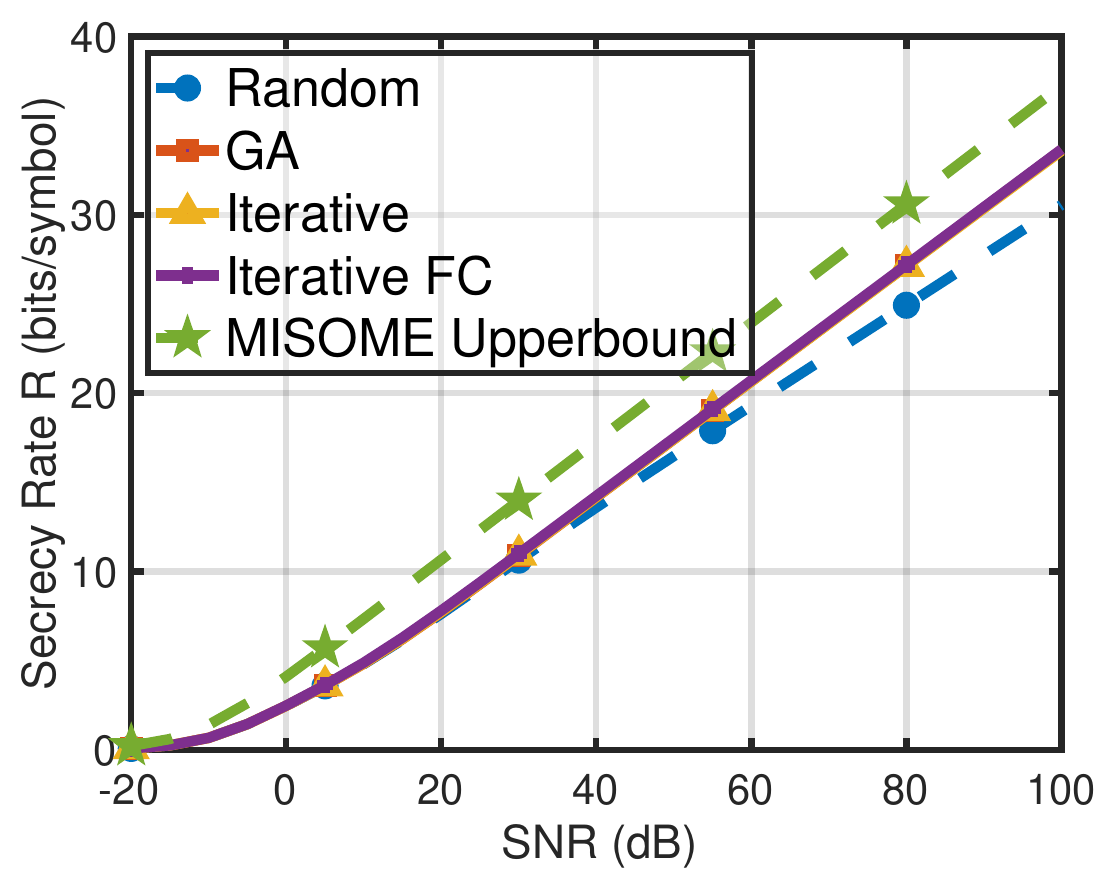}
    }
    \hfill
    \subfloat[$N_e = 5$, Rayleigh i.i.d. channel \label{fig:secrecy_N=20_Ne=5-Rayleigh}]{%
        \includegraphics[width=0.485\linewidth]{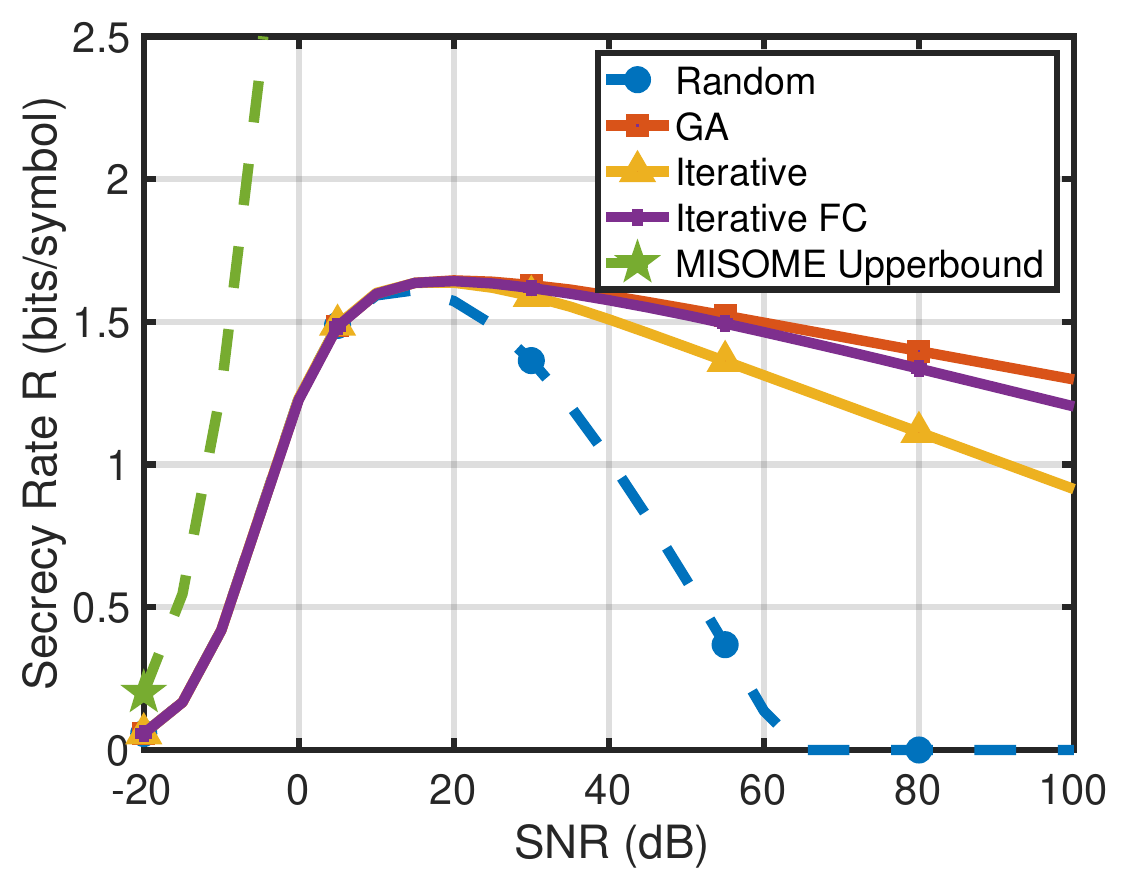}
    }
    \\
  \subfloat[$N_e = 1$, Geometric channel\label{fig:secrecy_N=20_Ne=1-geometric}]{%
            \includegraphics[width=0.485\linewidth]{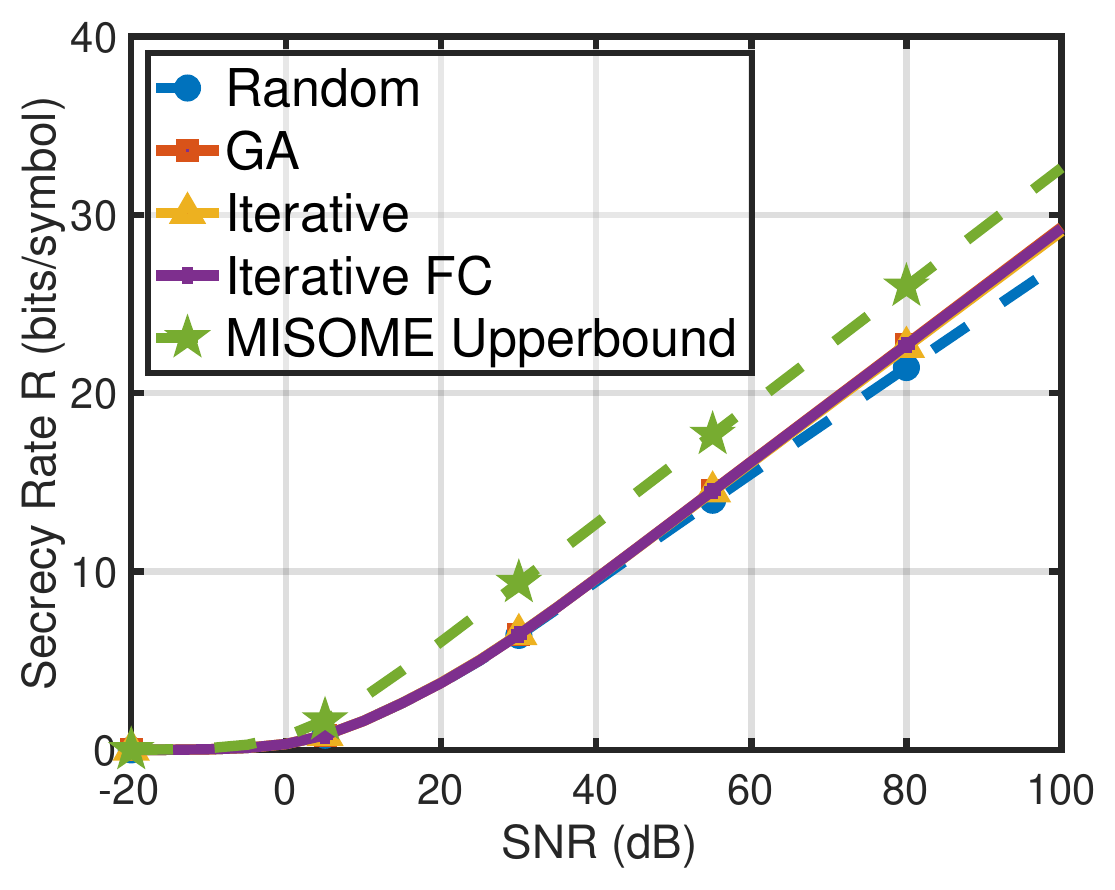}
    }
    \hfill
  \subfloat[$N_e = 5$, Geometric channel\label{fig:secrecy_N=20_Ne=5-geometric}]{%
            \includegraphics[width=0.485\linewidth]{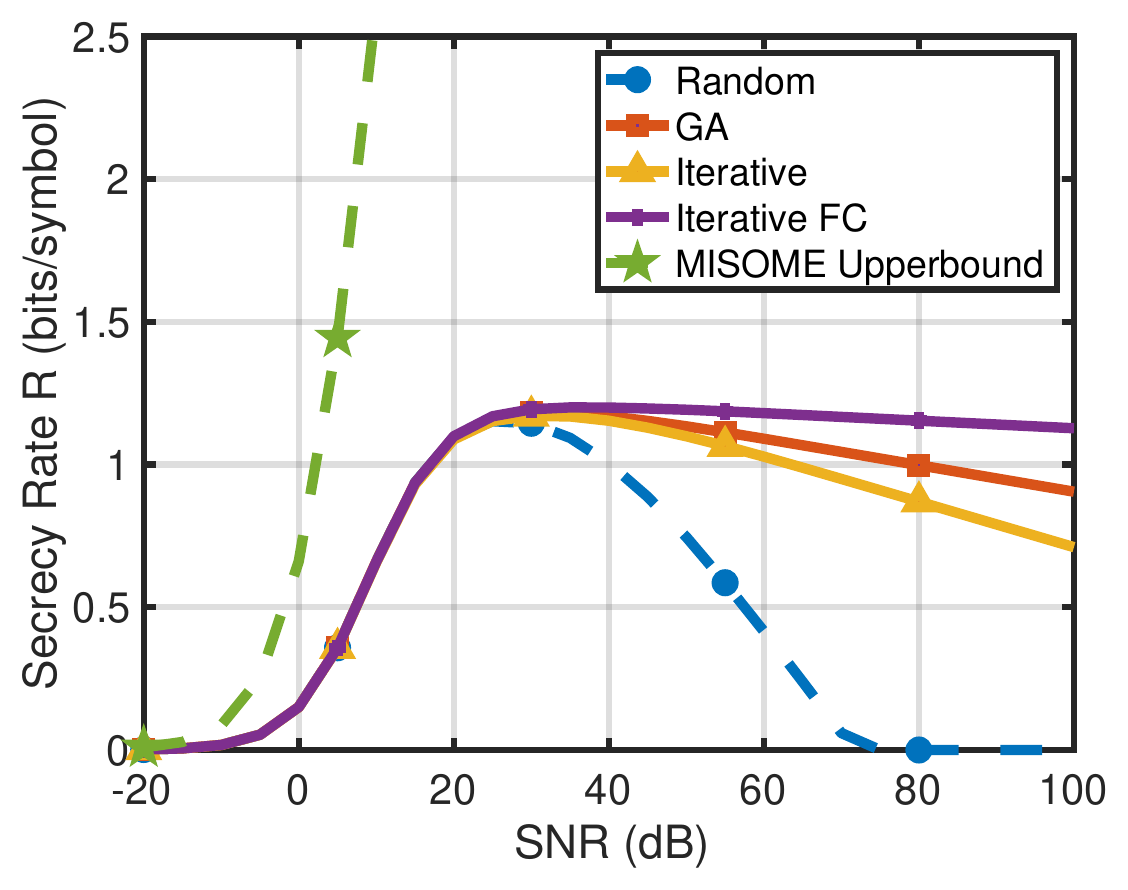}
    }
    \caption{Secrecy rates when $N = 20$ for the i.i.d. Rayleigh fading channel and the geometric channel with $L = 10$ paths.}
  \label{fig:secrecy-rate-N=20}
\end{figure}

\begin{figure}
    \centering
    \subfloat[$N_e = 10$, Rayleigh i.i.d. channel\label{fig:secrecy_N=30_Ne=10-Rayleigh}]{%
        \includegraphics[width=0.485\linewidth]{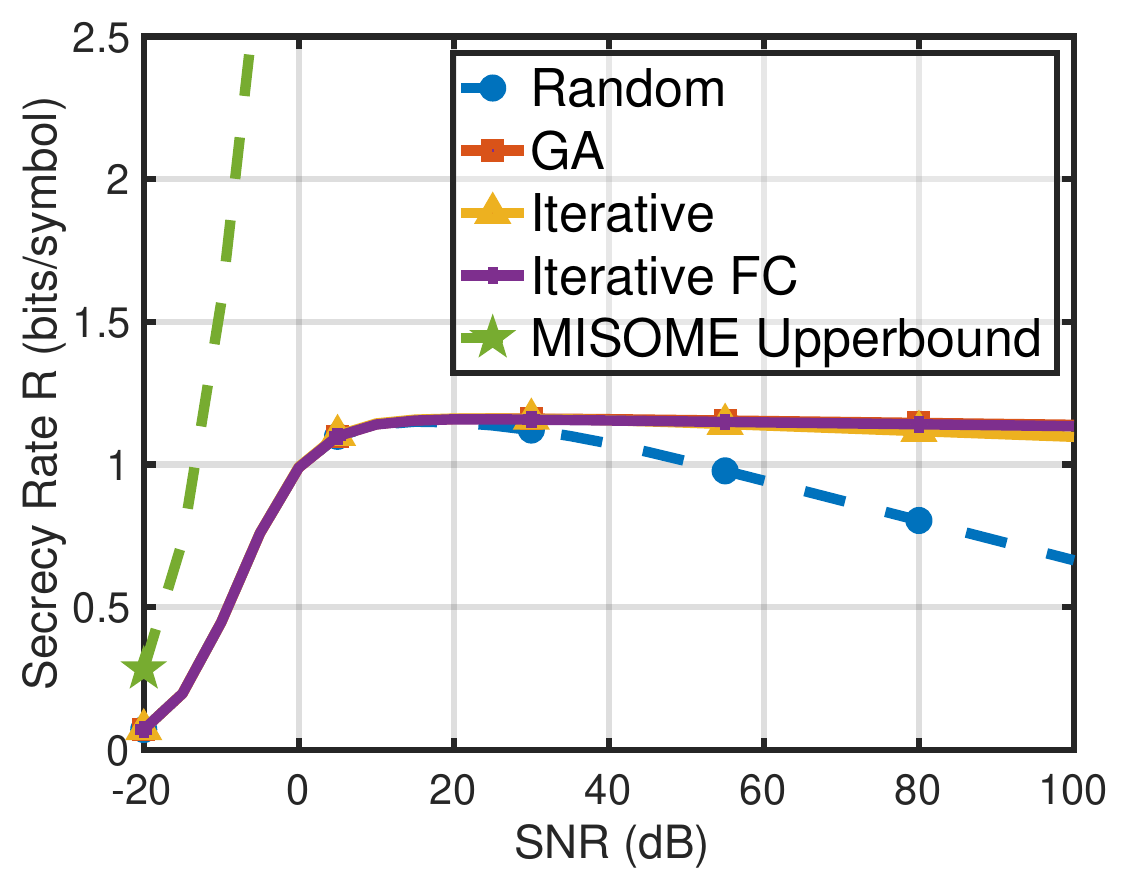}
    }
    \hfill
    \subfloat[$N_e = 10$, Geometric channel\label{fig:secrecy_N=30_Ne=10-geometric}]{%
        \includegraphics[width=0.485\linewidth]{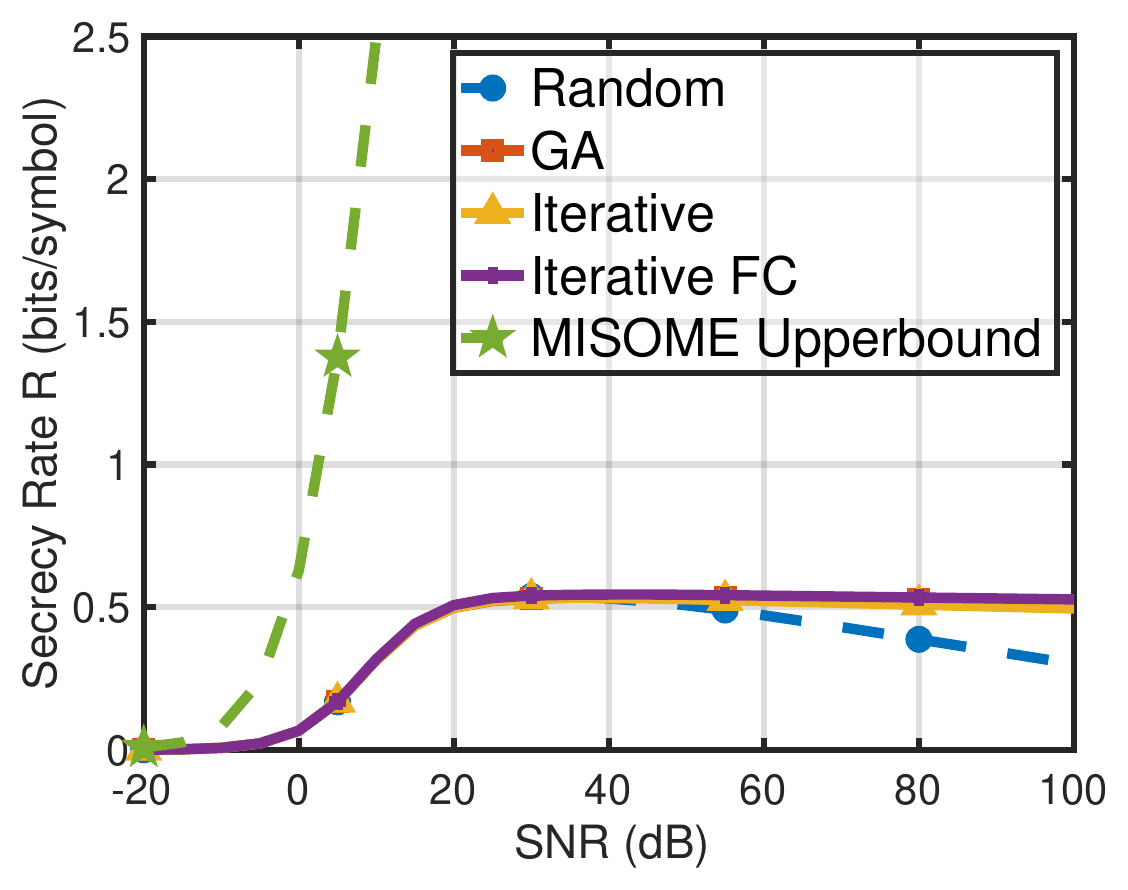}
    }
    \caption{Secrecy rates when $N = 30$ for the i.i.d. Rayleigh fading channel and the geometric channel with $L = 10$ paths. The secrecy rates of Iterative, Iterative FC, and GA closely overlap, indicating their similar performance.}
    \label{fig:secrecy-rate-N=30}
\end{figure}

Numerical results of the secrecy rate for the Rayleigh channel model are shown in Figure \ref{fig:secrecy_N=20_Ne=1-Rayleigh}, Figure \ref{fig:secrecy_N=20_Ne=5-Rayleigh}, and Figure \ref{fig:secrecy_N=30_Ne=10-Rayleigh} for the cases $(N, N_e) = (20, 1)$, $(20, 5)$, and $(30, 10)$. In addition to the 4 partition schemes, we plot the MISOME upper bound \cite{khisti_wornell_2010} for comparison.


We note that the secrecy rate serves as an indirect measure of the outage probability. 
Therefore, the pattern of outage probability in the proposed partition algorithms also extends to the secrecy rate. 
Our observations at $N = 20$ indicate that the more intelligent partition algorithms (iterative, iterative FC, and GA) outperform the baseline random partition algorithm. 
This difference becomes particularly pronounced at high SNR regions with a large number of eavesdropper antennas, as seen in Figure \ref{fig:secrecy_N=20_Ne=5-Rayleigh} ($Ne = 5$) and Figure \ref{fig:secrecy_N=30_Ne=10-Rayleigh} ($Ne = 10$).
In both plots, the secrecy rate decays when the SNR increase above $20$ dB as a result of increasing noise power at the legitimate receivers.
The decay can be mitigated by i) increasing $N$ (See the $N = 30$ case in Figure \ref{fig:secrecy_N=30_Ne=10-Rayleigh}), or ii) using a partition algorithm with a lower outage probability. Figures \ref{fig:secrecy_N=20_Ne=5-Rayleigh} and \ref{fig:secrecy_N=30_Ne=10-Rayleigh} show that the iterative and genetic algorithms do not decay as much as the random partition scheme does.

We also plot the geometric channel model with $L = 10$ in Figure \ref{fig:secrecy_N=20_Ne=1-geometric}, Figure \ref{fig:secrecy_N=20_Ne=5-geometric}, and Figure \ref{fig:secrecy_N=30_Ne=10-geometric}. In general, the patterns observed in the Rayleigh channel are also evident in the geometric channel.
Although we observed lower outage probabilities in the geometric channel, it yields a lower secrecy rate (and MISOME upper bound) than the Rayleigh i.i.d. channel. 
This discrepancy may be attributed to the non-independence of channel gains, which can be predicted more easily.

\subsection{Algorithm runtime}
Table \ref{tab:algs-runtime-mstar-rayleigh} compares the time complexity and simulated runtime of the three proposed algorithms for $N=20$ and $30$.  
We computed the averaged runtime results of the algorithms over $1000$ realizations of the i.i.d. Rayleigh fading channel.
While the genetic algorithm appears to be the fastest from an asymptotic perspective (scaling linearly with $N$), direct comparison with the other three algorithms is not straightforward due to significant constant factors affecting its runtime. The constant factors include the population size $P$ and the number of generations $G$, which are specific to the genetic algorithm. 
Although the iterative algorithms scale with $N^2$, the actual runtime is approximately 0.01 of the GA.
In summary, the random partition algorithm requires the shortest computation time, followed by the iterative algorithms, while the genetic algorithm requires the longest runtime.



\begin{table}[h]
    \caption{Complexity and simulated runtime for the i.i.d. Rayleigh fading channel.}
    \label{tab:algs-runtime-mstar-rayleigh}
    \resizebox{1\linewidth}{!}{
    \centering
    \begin{tabular}{c|c|c|c}
        \cmidrule{1-4}
        Algorithm & Complexity & Runtime ($N = 20$) & Runtime ($N = 30$) \\
        
        \cmidrule{1-4}
        Random & $\mathcal{O}(N)$ & 1.17 ms & 1.70 ms\\
        Iterative & $\mathcal{O}(K(N \log N + N^2))$ & 1.27 ms & 1.83 ms \\
        Iterative FC & $\mathcal{O}(K(N \log N + N^2))$ & 1.21 ms & 2.07 ms \\
        GA & $\mathcal{O}(G P N)$ & 81.11 ms & 216.21 ms \\

        \cmidrule{1-4}
    \end{tabular}
    }
    \label{tab:results}
\end{table}

\section{Conclusion}
\label{sec:conclusion}

We have proposed a novel successive partition zero-forcing scheme to solve the multi-user zero-forcing problem. 
This reduces the problem to optimizing channel partitioning to minimize the outage probability.
We analyzed the case of $K=2$ users and present three partition algorithms: random partition, iterative partition, and genetic partition. 
We showed theoretically that the random partition is capable of achieving arbitrarily low outage probability with sufficiently many transmit antennas.
The iterative and genetic partition algorithms are more complex, and exhibit substantial secrecy rates gain compared to the random partition in three scenarios: high SNR, small number of transmit antennas, and large number of eavesdropper antennas.



\section{Appendix}
\label{sec:appendix}
\subsection{Generalization of SPZF for the case of $K \geq 2$} \label{app:spzf-generalization}
The SPZF can be easily generalized to arbitrary $K$. There will be $K$ successive zero-forcing steps; in the $k$-th step, we construct the partition matrix $B_k$ based on the $k$-th user channel. Let the reduced channel vector of the $k$-th user have $m_k$ terms ($k = [1, K]$), then the partition matrix $\mathbf{B_k} \in \{0, 1\}^{m_{k+1} \times m_k}$ forms $m_{k+1}$ partition sets, i.e. maps a vector from $\mathbb{C}^{m_{k}}$ to $\mathbb{C}^{m_{k+1}}$, where the following inequality of the number of terms at each step is required:
\begin{equation}
    3 \leq m_{k+1} \leq \frac{m_k}{3}.
    \label{eq:m-inequality}
\end{equation}
The lower bound ensures a minimum of $3$ terms to form a polygon in the $(k+1)$-th step.  
The upper bound inequality ensures a minimum of $3$ terms in each of the $m_{k+1}$ partitions in the $k$-th step.

\par The inequality also imposes a scaling law for the minimum $N$ required for $K$ users. It is natural to denote $m_1 = N$ as the number of the terms in the first user channel $\mathbf{h_1}$. In the last step (i.e. the $K$-th step), we require $m_K \geq 3$. Applying the lower bound in Equation \ref{eq:m-inequality} recursively yields $m_1 = N \geq 3^K$.

\par The SPZF algorithm for general $K$ is given in Algorithm \ref{alg:successive-pzf-alg-general-K}.
\begin{algorithm}
    \caption{Successive zero-forcing algorithm for general $K$}\label{alg:successive-pzf-alg-general-K}
    \begin{algorithmic}[1]
        \Require $\mathbf{h_1}$, $\mathbf{h_2}, \dots, \mathbf{h_K}$
        \For{$k = 1, \dots, K-1$}
        \State Construct partition matrix $\mathbf{B_k}: \mathbb{C}^{m_k} \to \mathbb{C}^{m_{k+1}}$
        \State ($\mathcal{B}(\mathbf{h_k}) = \mathcal{B}_1, \dots, \mathcal{B}_{m_{k+1}}$ are the various partition sets.)
        \State $\underline{\phi}_K \gets \textrm{polygon\_solver}(\mathcal{B}(\mathbf{h_k}))$
        \State Apply $\mathbf{B_k} \diag{e^{j \underline{\phi}_K}}$ to all vectors $\mathbf{h_1}, \dots, \mathbf{h_k}$
        \EndFor
        \State $\mathbf{B_K} = [1, \dots, 1] \in {1}^{1 \times m_{K-1}}$
        \State $\underline{\phi}_K \gets \textrm{polygon\_solver}(\{ h_{Ki}\})$
        \State $\mathbf{w} \gets \prod_{k=1}^{K} \diag(e^{j \underline{\phi}_k}) \mathbf{B_k}^T$
        \Ensure $\mathbf{w}$
    \end{algorithmic}
\end{algorithm}

\subsection{Proof of Proposition \ref{prop:reduced-ch-vect-iid}} \label{app:reduced-ch-vect-iid}
We first show that for fixed $\mathbf{B_1} \diag{(e^{j \underline{\phi}_1})}$, $\mathbf{y} = \mathbf{B_1} \diag{(e^{j \underline{\phi}_1})} \mathbf{h_2}$ is jointly Gaussian. This is immediate from the fact (c.f. \cite[Theorem 2]{gallager2008circularly}) that a necessary and sufficient condition for a random vector to be a circularly-symmetric jointly Gaussian random vector is that it has the form $\mathbf{z} = \mathbf{A}\mathbf{w}$, where $\mathbf{w}$ is i.i.d. complex Gaussian and $\mathbf{A}$ is an arbitrary complex matrix.  

It thus only remains to show that the elements of $\mathbf{y}$ are jointly independent. For jointly Gaussian vectors, a necessary and sufficient condition is that the covariance matrix is a diagonal matrix. 
\begin{equation}
    \begin{aligned}
        \mathbf{K_y}   &= \Exp[\mathbf{y} \cdot \mathbf{y}^H] \\
        &= \Exp[ \mathbf{B_1} \diag{(e^{j \underline{\phi}_1})} \mathbf{h_2} \cdot (\mathbf{B_1} \diag{(e^{j \underline{\phi}_1})} \mathbf{h_2})^H] \\
        &= \mathbf{B_1} \diag{(e^{j \underline{\phi}_1})} \cdot \Exp[\mathbf{h_2} \mathbf{h_2}^H] \cdot (\mathbf{B_1} \diag{(e^{j \underline{\phi}_1})})^H \\
        &\overset{\text{(i)}}{=} \mathbf{B_1} \diag{(e^{j \underline{\phi}_1})} \cdot \sigma^2 \mathbf{I_N} \cdot (\mathbf{B_1} \diag{(e^{j \underline{\phi}_1})})^H \\
        &\overset{\text{(ii)}}{=} k \sigma^2 \mathbf{I_m},
    \end{aligned}
    \label{eq:y-covariance-matrix}
\end{equation}
where (i) follows from the fact that $\mathbf{h_2}$ has i.i.d. complex Gaussian elements.
(ii) follows from the observation that $\mathbf{B_1 B_1^H} = k \mathbf{I_m}$. 
To see this, let the rows of $\mathbf{B_1}$ be $b_1, \dots, b_m$. Note that $\braket{b_i, b_j} = | \mathcal{B}_{i} \cap \mathcal{B}_j |$.
Hence, $b_{ii} = \braket{b_i, b_i} = | \mathcal{B}_i | = k$. 
The off-diagonals have the form $\braket{b_i, b_j}$, which is $0$ since all partition sets are disjoint. $\qed$

\subsection{Proof of Proposition \ref{prop:pr-e2-e1c}} \label{app:pr-e2-e1c}
Here we implicitly condition all probabilities to partition sets within $\mathscr{B}_\text{FC}$. Let $\mathcal{D} = \{\bar{\mathbf{B}} = \mathbf{B_1}e^{j \diag{\underline{\phi}_1}} \,|\, \mathbf{B_1} \in \mathscr{B}_\text{FC}, \,  \underline{\phi}_1 \in [0, 2\pi] ^N\}$.
\begin{equation}
    \begin{aligned}
        \Pr(\varepsilon_2|\varepsilon_1^\mathsf{c}) 
        & \overset{\text{(i)}}{=} \int_{\mathcal{D}}  \Pr(\varepsilon_2|\varepsilon_1^\mathsf{c}, \bar{\mathbf{B}}= \mathbf{B} ) \Pr(\bar{\mathbf{B}} = \mathbf{B}) d \mathbf{B} \\
        & \overset{\text{(ii)}}{=} \int_{\mathcal{D}}  \Pr(\varepsilon_2| \bar{\mathbf{B}}= \mathbf{B} ) \Pr(\bar{\mathbf{B}} = \mathbf{B}) d \mathbf{B} \\
        & \overset{\text{(iii)}}{=} \int_{\mathcal{D}}  f_\text{Ray}(m)  \Pr(\bar{\mathbf{B}} = \mathbf{B}) d \mathbf{B} \\
        &= f_\text{Ray}(m),
    \end{aligned}
\end{equation}
where (i) follows from the total probability law;
(ii) follows from the observation that conditioned on a fixed $\bar{\mathbf{B}}$, $\mathbf{y}$ is only a function of $\mathbf{h_2}$. Hence, $\varepsilon_2 = \{\mathbf{h_2}: \textrm{dist} (\{|y_l|\}_{l=1}^m) > 0 \}$ is independent of $\mathbf{h_1}$ and consequently independent of $\varepsilon_1$;
(iii) follows from Definition \ref{defn:channel-error-funct} and the fact that $\{|y_l|\}_{l=1}^m$ are i.i.d. jointly Gaussian from Proposition \ref{prop:reduced-ch-vect-iid}. $\qed$
\subsection{Proof of Proposition \ref{prop:loss-to-prob}} \label{app:loss-to-prob}
    Recall from Definition \ref{defn:error-prob}, $\varepsilon_1$ is the probability that any partition $\mathcal{B}_l$ from the partition set $\mathcal{B}_1, \mathcal{B}_2, \dots, \mathcal{B}_m$ fails the polygon inequality. 
    \begin{equation}
        \begin{aligned}
            \Pr(\varepsilon_1)  & = \Pr ( \{ \max_{l = 1, \dots, m} \textrm{dist}(\mathcal{B}_l) > 0 \} )  \\
            & = \int_{\mathbf{h_1}} \mathbbm{1} [\max_{l = 1, \dots, m} \textrm{dist}(\mathcal{B}_l) > 0] p(\mathbf{h}) d\mathbf{h} \\
            & = \int_{\mathbf{h_1}}  \mathcal{H} [\max_{l = 1, \dots, m} \textrm{dist}(\mathcal{B}_l)] p(\mathbf{h}) d\mathbf{h} \\
            & = \int_{\mathbf{h_1}}  E(\mathcal{B}_1, \mathcal{B}_2, \dots, \mathcal{B}_m)  p(\mathbf{h}) d\mathbf{h} \\
            & = \Exp{[E]},
        \end{aligned}
    \end{equation}
    where $p(\cdot)$ is the probability distribution function for $\mathbf{h_1}$. $\qed$

\subsection{Proof of Proposition \ref{prop:min-e-min-pr}} \label{app:min-e-min-pr}
We show the proof in two steps. 
\begin{enumerate}
    \item $e_{k+1} \leq e_k \implies \Exp[E_k] \leq \Exp[E_{k+1}]$.
    \item $\Exp[E_k] \leq \Exp[E_{k+1}] \implies$ $\Pr(\varepsilon_1)$ is minimized to a local minimum.
\end{enumerate}
(1) follows from the observation $(e_{k+1} \leq e_k) \implies (\mathcal{H} (e_{k+1}) \leq \mathcal{H} (e_k))$ . Thus,
\begin{equation}
    \begin{aligned}
        \Exp{[E_{k+1}]}     &= \int_{\mathbf{h_1}} \mathcal{H} (e_{k+1}) p(\mathbf{h}) d\mathbf{h} \\
        & \leq \int_{\mathbf{h_1}} \mathcal{H} (e_{k}) p(\mathbf{h}) d\mathbf{h} \\
        &= \Exp{[E_k]},
    \end{aligned}
\end{equation}
where $p(\cdot)$ is the probability distribution function of $\mathbf{h_1}$. (2) follows from $\Pr(\varepsilon_1) = \Exp{[\lim\limits_{k \to \infty} E_k]}$. $\qed$
    \subsection{Proof of Proposition \ref{prop:random-part-err-limit}} \label{app:random-part-err-limit}
    To emphasize the fixed number of partition sets, we denote $m = m_0$. We approximate $(1 - f_\text{Ray}(\frac{N}{m_0}))$ using Equation \ref{eq:fm-error}:
        \begin{equation}
                \begin{aligned}
                    \Pr [\mathtt{outage}]
                    & = 1 - ( 1- f_\text{Ray}(\frac{N}{m_0})) ^ {m_0}  (1 - f_\text{Ray}(m_0)) \\
                    & \approx 1 - ( 1- \frac{N}{m_0} e^{-\frac{\pi}{16} (\frac{N}{m_0})^2})^{m_0} (1 - f_\text{Ray}(m_0)).
                \end{aligned}
            \label{eq:random-outage-approx1}
        \end{equation}

        Note that $\frac{N}{m_0} e^{-\frac{\pi}{16} (\frac{N}{m_0})^2} \to 0$ in the limit that $N \to \infty$. For small $x$, $(1 - x)^{m_0} \approx 1 - m_0 x$. With $ x = \frac{N}{m_0} e^{-\frac{\pi}{16} (\frac{N}{m_0})^2}$): 
        \begin{equation}
            \begin{aligned}
                & \lim\limits_{N \to \infty} \Pr [\mathtt{outage}] \\
                & = \lim_{N \to \infty} 1 - (1 - m_0 \frac{N}{m_0} e^{- \frac{\pi}{16} (\frac{N}{m_0})^2})(1 - f_\text{Ray}(m_0)) \\
                & = \lim_{N \to \infty} 1 - (1 - N e^{- \frac{\pi}{16} (\frac{N}{m_0})^2})(1 - f_\text{Ray}(m_0))\\
                & = \lim_{N \to \infty} 1 - (1 - f_\text{Ray}(m_0)) \\
                & = \lim_{N \to \infty} f_\text{Ray}(m_0),
            \end{aligned}
        \end{equation}
        which completes the proof. $\qed$

\subsection{Iterative partition algorithm with fixed cardinality (FC) constraint} \label{app:iterative-fc-pseudocode}
        \begin{algorithm}
            \caption{Iterative partition algorithm with a fixed cardinality constraint. The differences from the unconstrained iterative partition algorithm are highlighted in \red{red}.}\label{alg:iter-part-fc}
            \begin{algorithmic}[1]
                \Require $\{h_i\}_{i=1}^N$
                \Require $m$ (number of partition sets)

                \State \textbf{Initialize:}
                \Indent
                \State m partition sets $\mathcal{B}_1, \mathcal{B}_2, \dots, \mathcal{B}_m$.
                \State For all $h_i$, randomly assign to one of $\mathcal{B}_1, \mathcal{B}_2, \dots, \mathcal{B}_m$.
                \EndIndent
                \While{ $E(\mathcal{B}_1, \mathcal{B}_2, \dots, \mathcal{B}_m) > 0$ or $E(\mathcal{B}_1, \mathcal{B}_2, \dots, \mathcal{B}_m)$ converges}
                \ForAll{$\mathcal{B}_l$ with $\textrm{dist}(\mathcal{B}_l) > 0$, in increasing order}
                \State \red{d = $\textrm{dist}(\mathcal{B}_l)$ + $\min(\mathcal{B}_l)$}
                \State \texttt{pocket\_d} $\gets d$

                \ForAll {$\{h_i\}$, $|h_i| > d,$ in increasing order}
                \State Let $h_i \in \mathcal{B}_s \neq \mathcal{B}_l$
                \If {$\textrm{dist}(\mathcal{B}_s \setminus \{h_i\})< $ \texttt{pocket\_d} }
                \State \red{\texttt{pocket\_d} $\gets \textrm{dist}(\mathcal{B}_s \setminus \{h_i\} \cup \min(\mathcal{B}_l))$}
                \State \texttt{pocket\_B} $\gets \mathcal{B}_s$
                \State \texttt{pocket\_hi} $\gets h_i$

                \If {  \texttt{pocket\_d} $ < 0$ }
                \State \textbf{break for}
                \EndIf
                \EndIf
                \EndFor

                \If {\texttt{pocket\_d} $ < d$}
                \State $\mathcal{B}_s \gets $ \texttt{pocket\_B}
                \State \red{$\mathcal{B}_s \gets \mathcal{B}_s \setminus \{\texttt{pocket\_hi}\} \cup \min(\mathcal{B}_l)$}
                \State \red{$\mathcal{B}_l \gets \mathcal{B}_l \cup \{\texttt{pocket\_hi}\} \setminus \min(\mathcal{B}_l)$}
                \EndIf

                \EndFor
                \EndWhile

                \Ensure $\mathcal{B}_1, \mathcal{B}_2, \dots, \mathcal{B}_m$
            \end{algorithmic}
        \end{algorithm}

\bibliographystyle{IEEEtran}
\bibliography{main}

\end{document}